\documentclass[12pt]{iopart}
\expandafter\let\csname equation*\endcsname\relax
\expandafter\let\csname endequation*\endcsname\relax
\usepackage{amsfonts} 
\usepackage{amsmath}
\usepackage{iopams}
\usepackage{amssymb}
\usepackage{mathtools}
\usepackage{cancel} 
\usepackage{bm}
\usepackage[usenames, dvipsnames]{color} 
\usepackage{dcolumn}
\usepackage{epic} 
\usepackage{epsfig}
\usepackage{feynmp}
\usepackage{graphicx}
\usepackage{grffile}
\usepackage[breaklinks,colorlinks = true,linkcolor = red,urlcolor=blue,citecolor=red]{hyperref}
\usepackage{mathrsfs}
\usepackage{soul}
\usepackage{subfigure}
\usepackage{wrapfig}
\usepackage{xy} 
\usepackage{xcolor}
\usepackage{amsmath,amssymb,amsfonts,amsthm}
\usepackage[ascii]{inputenc}
\usepackage{makecell}
\allowdisplaybreaks

\graphicspath{{images/}{../Figures}}

\begin{document}

\title{Harmonically confined long-ranged interacting gas in the presence of a hard wall}

\author{Jitendra Kethepalli, Manas Kulkarni,  Anupam Kundu}
\address{International Centre for Theoretical Sciences, Tata Institute of Fundamental Research, Bengaluru -- 560089, India}
\author{Satya N. Majumdar}
\address{LPTMS, CNRS, Univ.  Paris-Sud,  Universit\'e Paris-Saclay,  91405 Orsay,  France}
\author{David Mukamel}
\address{Department of Physics of Complex Systems, Weizmann Institute of Science, Rehovot 7610001, Israel}
\author{Gr\'egory Schehr}
\address{Sorbonne Universit\'e, Laboratoire de Physique Th\'eorique et Hautes Energies, CNRS UMR 7589, 4 Place Jussieu, 75252 Paris Cedex 05, France}



\begin{abstract}
In this paper, we compute exactly the average density of a harmonically confined Riesz gas of $N$ particles for large $N$ in the presence of a hard wall. 
In this Riesz gas, the particles repel each other via a pairwise interaction that behaves as $|x_i - x_j|^{-k}$ for $k>-2$, with $x_i$ denoting the position of the $i^{\rm th}$ particle. This density can be classified into three different regimes of $k$. For $k \geq 1$, where the interactions are effectively short-ranged, the appropriately scaled density has a finite support over $[-l_k(w),w]$ where $w$ is the scaled position of the wall. While the density vanishes at the left edge of the support, it approaches a nonzero constant at the right edge $w$. For $-1<k<1$, where the interactions are weakly long-ranged, we find that the scaled density is again supported over $[-l_k(w),w]$. While it still vanishes at the left edge of the support, it diverges at the right edge $w$ algebraically with an exponent $(k-1)/2$. For $-2<k< -1$, the interactions are strongly long-ranged that leads to a rather exotic density profile with an extended bulk part and a delta-peak at the wall, separated by a hole in between. Exactly at $k=-1$ the hole disappears. For $-2<k< -1$, we find an interesting first-order phase transition when the scaled position of the wall decreases through a critical value $w=w^*(k)$. For $w<w^*(k)$, the density is a pure delta-peak located at the wall. The amplitude of the delta-peak plays the role of an order parameter which jumps to the value $1$ as $w$ is decreased through $w^*(k)$.
Our analytical results are in very good agreement with our Monte-Carlo simulations. 
\end{abstract}
\date{\today}
\maketitle

\section{Introduction}
\label{intro} 
Long-ranged interacting particle systems are ubiquitous in nature and have been a subject of great interest in physics and mathematics. Especially  systems with pairwise repulsive interactions are important due to their appearances in many contexts such as in the physics of cold atoms~\cite{brown2003rotational, chalony2013long, zhang2017observation}, random matrix theory (RMT)~\cite{mehta2004random,  forrester2010log}, integrable models~\cite{olshanetsky1981classical, polychronakos2006physics, kulkarni2017emergence} , gravitational systems~\cite{padmanabhan1990gravity, marcos2013collisional}, hydrodynamics~\cite{miller1990statistical, robert1990etats, kulkarni2012hydrodynamics}, synchronisation~\cite{rakshit2020synchronization} and plasma physics~\cite{elskens2003microscopic} to name a few. In particular, systems confined in an external potential and with pairwise power law interactions have recently drawn a lot of interest. Specific examples of such systems with long-range interactions include, e.g., one-dimensional one-component plasma ($1d$OCP)~\cite{lenard61, prager62,baxter1993statistical, dhar2017exact, dhar2018extreme}, Dyson's log gas~\cite{mehta2004random,forrester2010log,dyson1962statistical1, dyson1962statistical2, dean2006large, dean2008extreme, Saffbook,majumdar2014top}, Calogero-Moser Systems~\cite{polychronakos2006physics, calogero1971solution, calogero1975exactly}, dipolar Bose gas~\cite{griesmaier2005bose, lu2011strongly}, ionic systems~\cite{brown2003rotational,zhang2017observation, yan2016exploring}, $3d$ Coulomb gas confined in one dimension ~\cite{dubin1997minimum} and Yukawa gas~\cite{cunden2018universality}. 
These systems belong to a broad class of power law models called Riesz gas composed of $N$ particles described by the following energy function  ~\cite{marcelriesz1938}
\begin{equation}\label{hamiltonian}
    E_k(\{x_i\}) = \sum_i^N \frac{x_i^2}{2} + \frac{J \, \text{sgn}(k)}{2} \sum_{i \neq j}^N |x_i-x_j|^{-k}, 
\end{equation}
where $x_i$'s are the positions of the particles and the exponent $k$ characterizes the range of the interaction with $\text{sgn}(k)$ ensuring the repulsive behaviour. The limit $k \to 0$ is a bit singular, where upon setting $J = 1/k$ the interaction term can be shown to be proportional to $-(1/2)\ln{|x_i-x_j|}$, which corresponds to Dyson's log gas in RMT~\cite{mehta2004random, forrester2010log}. In the rest of the paper, to keep the notations light, we will set $J=1$, keeping in mind that the $k \to 0$ limit is special. Several properties of the Riesz gas for specific values of $k$ have been studied both in the physics and in the mathematics literature. For example, as mentioned above, $k\to0$ corresponds to the Dyson's log gas appearing in random matrix theory~\cite{mehta2004random, forrester2010log, dean2006large,dean2008extreme}, $k =-1$ corresponds to the one-dimensional one component plasma ($1d$OCP)~\cite{ forrester2010log,lenard61, prager62,baxter1993statistical,dhar2017exact, dhar2018extreme} and $k=2$ corresponds to the well known Calogero-Moser model which is an interacting integrable system~\cite{polychronakos2006physics, calogero1971solution, calogero1975exactly,  agarwal2019some}. Recently, for the Riesz gas with a confining harmonic potential in Eq. (\ref{hamiltonian}), the average density in the thermal equilibrium for large $N$ has been computed exactly for all $k>-2$ \cite{agarwal2019harmonically}. This calculation was extended recently to finite-range and other non-harmonic confining potentials~\cite{kumar2020particles}.

Due to the long-ranged nature of the interaction, the equilibrium properties of the Riesz gas provide an interesting example of the thermodynamics of non extensive systems~\cite{campa2009statistical,review_david}. In higher dimensions, the Riesz gas for large $k$ is related to the sphere packing problem which is  a classic optimization problem~\cite{saff1997distributing}. For the Riesz gas in thermal equilibrium at inverse temperature $\beta$, the joint probability distribution function of the positions of the particles is described by the Boltzmann distribution
\begin{eqnarray} \label{P_Boltz}
P_k(\{x_i\}) =\frac{1}{Z_k(\beta)} e^{-\beta E_k\left(\{x_i\}\right)} \;,
\end{eqnarray}
where the partition function $Z_k(\beta)= \int dx_1 dx_2 \ldots dx_N\, e^{-\beta E_k(\{x_i\})}$ normalises this probability distribution.

One of the most natural and basic questions is: what is the average density of the Riesz gas in thermal equilibrium, in the limit of large $N$? This question has been studied extensively in one dimension for $k > -2$ and is relevant in many different contexts~\cite{agarwal2019harmonically,serfaty2015, leble2018}. It turns out that for $k< - 2$, the system is unstable, in the sense that even in the ground state the particles fly away to $\pm \infty$. Recent exact results for large $N$ in Ref.~\cite{agarwal2019harmonically} have shown that the shape and the scale of the average density profile depend crucially on the value of $k$. For any $k$ and $\beta = {\cal O}(1)$, the average density in the large $N$ limit has a finite support and is described by the following scaling form~\cite{agarwal2019harmonically}
\begin{equation}\label{unconstrained-density}
   \langle \rho_N(x)  \rangle = \frac{1}{N^{\alpha_k}} \rho_{k, {\rm uc}}^{\rm{ *}}\left(\frac{x}{N^{\alpha_k}}\right),
\end{equation}
where the scaling function $\rho_{k, {\rm uc}}^{\rm{*}}\left(y\right)$ is of the form 
\begin{equation} \label{rho_uc}
    \rho_{k, {\rm uc}}^{\rm{*}}(y) = \frac{\left(1-(y/l_k^{\rm uc})^2\right)^{\gamma_k}
}{2^{\gamma_k} l_k^{\rm uc}B\left(\gamma_k+1, \gamma_k+1\right)},~~\text{for}~ -l_k^{\rm uc} \leq y \leq l_k^{\rm uc},
\end{equation}
with $B(x,y)$ being the standard Beta function and
\begin{equation}
\gamma_k = 
    \begin{cases}
 \frac{1}{k}, &\text{for}~k\geq1\\
      \frac{k+1}{2}, & \text{for}~-2<k<1
    \end{cases}
    \label{gamma_k} \;.
\end{equation} 
The exponent $\alpha_k$ in Eq. (\ref{unconstrained-density}) is given by
\begin{equation} \label{alpha_k}
    \alpha_k = 
    \begin{cases}
        \frac{k}{k+2}, &\text{for}~~~~~k \geq 1\\
        \frac{1}{k+2}, &\text{for}~ -2< k \leq 1\;. \\
    \end{cases}
\end{equation}
Thus the scaled density is supported over $[-l_k^{\rm uc}, +l_k^{\rm uc}]$ where $l_k^{\rm uc}$ depends on the model parameters and is given explicitly by~\cite{agarwal2019harmonically}
\begin{align} 
l_k^{\rm uc} = \frac{1}{2}\left(A_k B(\gamma_k+1, \gamma_k+1)\right)^{-\alpha_k}~~ \text{with}
~~ A_k = 
  \begin{cases}
  \left(2 \left(k+1\right) \zeta\left(k\right)\right)^{-\gamma_k}, & \text{for}~k>1\\
  \frac{1}{4 }, & \text{for}~k=1\\
  \frac{\sin\left(\pi \gamma_k\right)}{2  \pi \gamma_k  |k|}, & \text{for}~k<1\;. \end{cases}
  \label{A_1k}
\end{align}
Here $\zeta (k) = \sum_{n=1}^{\infty} 1/n^k$ is the Riemann Zeta function. In the above description, and in the rest of the paper, the superscript ``${\rm uc}$'' refers to unconstrained Riesz gas. Similarly, throughout the paper, we will reserve the superscript ``$*$'' to denote the saddle-point solutions in the large $N$, as will become clearer later. 
\begin{figure}
 \centering
        \includegraphics[scale=0.5]{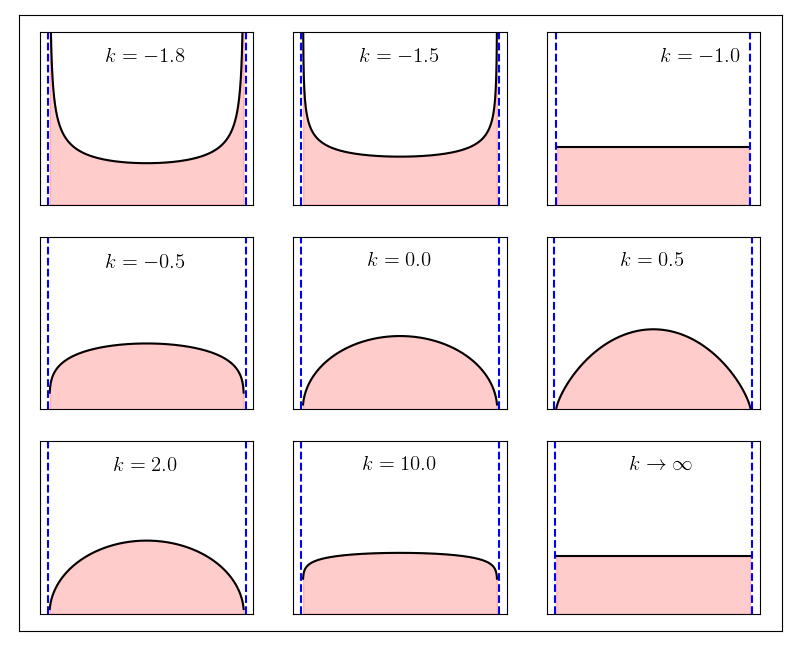}
    \caption{Scaled average density of the unconstrained gas $\rho_{k, {\rm uc}}^{\rm{*}}(y)$ in Eq. (\ref{rho_uc}) vs. $y$ for different values of $k>-2$. The three rows correspond respectively to (i) $k\geq 1$ (third row), (ii) $-1< k < 1$ (second row) and (iii) $-2 < k \leq -1$ (first row). In (i) the density has a dome shape, in (ii) also a dome shape and in (iii) it has a U-shape. At $k=-1$ and $k \to +\infty$ the density is flat. The blue dashed vertical lines indicate the edges of the support of the density profile. 
    }\label{fig:allkdens}
\end{figure}

\begin{figure}[t]
\centering
\includegraphics[width = 0.9\linewidth]{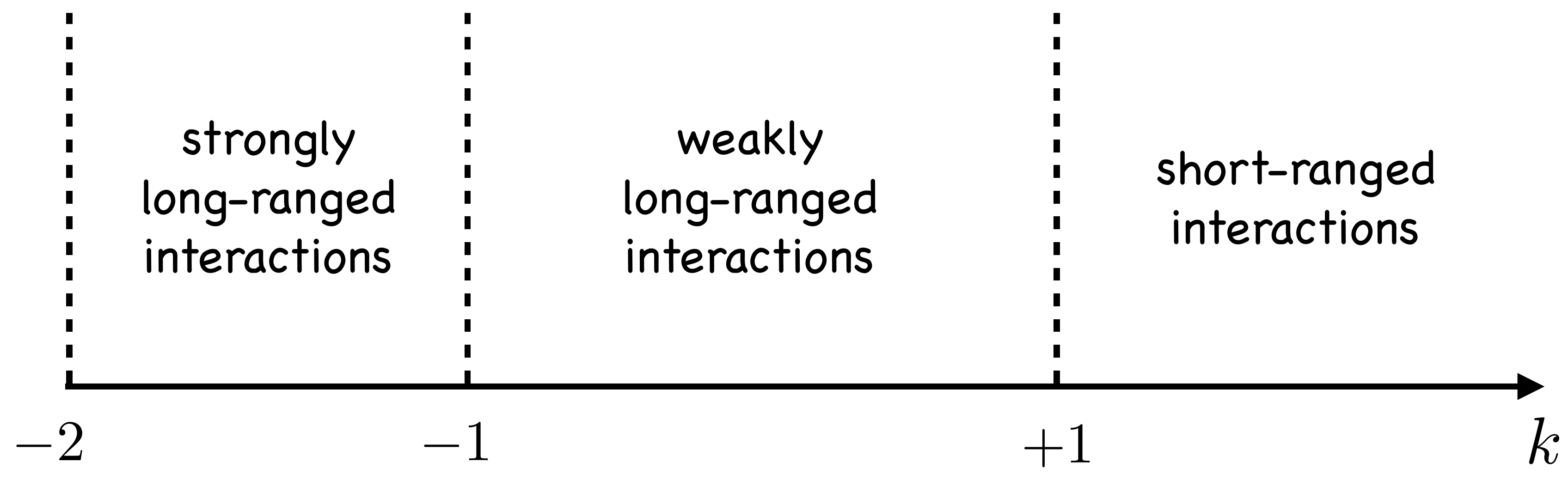}
\caption{The exponent $k$ characterizes the range of pairwise repulsive interactions between particles. For $k\geq 1$ the interaction is effectively short-ranged. For $-1<k<1$, the interaction, though long-ranged, is effectively weak and does not qualitatively change the shape of the density profile compared to the short-ranged case. For $-2<k\leq -1$, the particles are subjected to strong long-range interactions, that change drastically the shape of the density profile.}\label{Fig_k}
\end{figure}

The average density profile in Eq. (\ref{rho_uc}) is plotted for different values of $k$ in Fig.~\ref{fig:allkdens}. 
The behavior of the density profile can be classified into three different regimes of $k$ depending on the range of the pairwise interactions (see Fig.~\ref{Fig_k}): 
\begin{itemize}
\item[(i)] {\it Regime 1 ($k \geq 1$): short-ranged interactions}. In this range the interaction falls off rather rapidly, as a function of the separation between two particles. Consequently, one can replace the pairwise long-ranged interaction by an effective short-ranged one. In this regime, the average density is dome-shaped with 
a maximum at the center and the density vanishes at the two edges of the support (see the third row of Fig. \ref{fig:allkdens}).
\item[(ii)]{\it Regime 2 ($-1<k<1$): weakly long-ranged interactions}. In this case, the long-ranged interaction can not be replaced by an effective short-ranged one as in regime 1 above. Nevertheless, the average density still remains dome-shaped, as in regime 1 (see the second row of Fig. \ref{fig:allkdens}). Hence we call this regime as ``weakly long-ranged''.  	
\item[(iii)]{\it Regime 3 $(-2<k \leq -1)$: strongly long-ranged interactions}. In this regime, the repulsive force between two particles (i.e., the derivative of the pairwise interaction potential) vanishes when they get closer to each other. However, at large distances, the force increases with separation $|r|$ as a power law $\sim |r|^{|k|-1}$, making this a ``strongly long-ranged'' system. This affects rather strongly the shape of the average density profile. In fact the density profile is now `U-shaped' where it diverges at the two edges (but still integrable) and has a minimum at the center of the support (see the first row of Fig. \ref{fig:allkdens}).   
\end{itemize}
The density is completely flat exactly at $k=-1$. Furthermore, the system undergoes a change of behaviour at $k=1$. This is also manifest in the $k$-dependence of the exponents $\alpha_k$ and $\gamma_k$ in Eqs. (\ref{alpha_k}) and (\ref{gamma_k}) where one sees a drastic change of behavior as $k$ crosses the value $k=1$. Note that the classification of the three regimes above is based on the shape of the density profile. This is somewhat different from the nomenclature (short-ranged/weakly long-ranged and strongly long-ranged) typically used in the literature on long-ranged interacting particle systems, where the classification is based on the thermodynamic behavior of the free energy \cite{review_david} .

The main goal of this paper is to study how the average density profile in the unconstrained Riesz gas gets modified in the presence of a wall at position $W$, such that the particles
are constrained to stay to the left of the wall. This is a naturally interesting question in any interacting particle system: how does the presence of a hard wall affect the collective properties of the system, such as the average density? Recent experimental progress have made ultra-cold gases an ideal platform to explore such collective behaviour. In many experiments involving ultra cold gases, appropriate barriers are introduced to create desired non-trivial density profiles in a controllable/tunable manner~\cite{kulkarni2012hydrodynamics, andrews1997observation, o2002observation, joseph2011observation, gaunt2013bose}. It is also an important intermediate step in the computations of extreme value statistics (EVS) in a strongly interacting system \cite{evs_review}. In EVS, one is interested in the distribution of the position of the rightmost particle, $x_{\max}$. Using the Bolzmann distribution in Eq.~(\ref{P_Boltz}), the cumulative distribution of $x_{\max}$, in thermal equilibrium at inverse temperature $\beta$, is given by \cite{majumdar2014top}
\begin{eqnarray}\label{EVS}
{\rm Prob.}(x_{\max} \leq W) = \frac{1}{Z_N(\beta)} \int_{-\infty}^W\,dx_1 \ldots  \int_{-\infty}^Wdx_N \; e^{- \beta E_k[\{x_i\}]} \;.
\end{eqnarray}
To evaluate this restricted partition function in the large $N$ limit using the saddle-point method, one needs to compute the density that optimises the multiple integral in Eq.~(\ref{EVS}). This optimal density, in the large $N$ limit, also coincides with the average density in the presence of a wall. Hence, computing the average density in the presence of a wall is the first step towards calculating the EVS in this strongly interacting system.

The average density profile of the constrained Riesz gas (i.e., in the presence of a wall at $W$) has in fact been computed, for large $N$, for two special values of $k$: namely $k \to 0$ limit (Dyson's log gas) and $k = -1$ ($1d$OCP). For the $k \to 0$ case,  it was shown in Refs.~\cite{dean2006large, dean2008extreme} that the constrained density profile satisfies the scaling form as given in Eq.~\eqref{unconstrained-density} and the scaling function is given by 
\begin{equation} \label{rho_loggas}
    \rho^*_0(y,w) =
    \begin{cases}
    \frac{1}{\pi} \sqrt{(l_0^{\rm uc)^2} - y^2}, &\text{  with  } -l_0^{\rm uc}\leq y \leq l_0^{\rm uc} ~~~~ \text{for} \quad w >l_0^{\rm uc} \;,\\
     & \\ 
    \frac{1}{2 \pi}\sqrt{\frac{y + l_0(w)}{w-y}} \left[w+ l_0(w) - 2y \right], &\text{  with  } -l_0(w)\leq y \leq w ~~ \text{for} \quad w<l_0^{\rm uc} \;,
    \end{cases}
\end{equation}
where $w=W/\sqrt{N}$, $l_0^{\rm uc}=\sqrt{2}$ and $l_0(w) = \frac{2\sqrt{w^2+6} - w}{3}$.  For $w > l_0^{\rm uc}$, the gas does not feel the presence of the wall at $w$ and the density is the same
as that of the unconstrained gas, i.e., the Wigner semi-circular form, as given in the first line of Eq. (\ref{rho_loggas}). In contrast, when $w<l_0^{\rm uc}$, the gas gets
pushed by the wall. This leads to a complete re-organization of the charges and the density gets drastically modified from the Wigner semi-circular law, as given in the second
line of Eq. (\ref{rho_loggas}), where the density vanishes at the left edge of the support $l_0(w)$, while it diverges as a square-root singularity $\sim \frac{1}{\sqrt{w-y}}$ at the right edge of the support located at $w$. This integrable divergence indicates an accumulation of charges at the wall when the gas is pushed \cite{majumdar2014top}.

However, the presence of the pushing wall affects the density for the $1d$OCP ($k=-1$) differently. It has been shown in Refs.~\cite{dhar2017exact, dhar2018extreme} that in this case the constrained density profile is given by
\begin{equation}\label{rho_CG}
    \rho^*_{-1}(y,w) =
    \begin{cases}
    \frac{1}{2 l_{-1}^{\rm uc}}, &\text{with} -l_{-1}^{\rm uc} \leq y \leq l_{-1}^{\rm uc} \quad \text{   for } \quad w>l_{-1}^{\rm uc}\\
    \frac{1}{2 l_{-1}^{\rm uc}} + \frac{l_{-1}^{\rm uc}-w}{2 l_{-1}^{\rm uc}} \delta(y-w)\,, &\text{with} -l_{-1}^{\rm uc}(w)\leq y \leq w ~~\text{for} ~~ -l_{-1}^{\rm uc}<w < l_{-1}^{\rm uc}\\
    \delta(w-y), &\text{  with  } y \leq w  \quad \quad \quad \quad \quad \text{ for }  \quad w < -l_{-1}^{\rm uc}
    \end{cases}
\end{equation}
where $l_{-1}^{\rm uc}=1$. As in the log gas, for $w > l_{-1}^{\rm uc}$, the gas does not feel the presence of the wall and the average density has the same flat profile as in the
unconstrained case. When $w <  l_{-1}^{\rm uc}$, the particles feel the presence of the wall, leading to a re-organisation of the particles, as in the log gas. However, how they get reorganised for $k=-1$ is drastically different from the $k \to 0$ case. For $k = -1$, the particles that get displaced by the wall get fully absorbed inside the wall, leading to a delta-function peak at $y=w$, that coexists with an undisturbed flat bulk to the left of $w$. This leads to the density in the second line of Eq. (\ref{rho_CG}). Finally, when $w <  -l_{-1}^{\rm uc}$, all the particles get absorbed in the wall, leading to a single delta-peak, as given in the third line of Eq.~(\ref{rho_CG}).

These two specifics cases of $k$ suggest that the shape of the density profile gets affected dramatically due to the presence of the wall. In this paper, we compute 
exactly, for all $k >-2$, the density profile of the constrained gas in the presence of a wall. In the next section we summarise our main results. The derivations of our results
are provided in Section~\ref{derivation}. Section ~\ref{conclusion} contains a summary and conclusions. Some details are relegated to the Appendix.

\section{Summary of the results}
\label{summary}

\begin{figure}[t]
 \centering
    \includegraphics[scale=0.5]{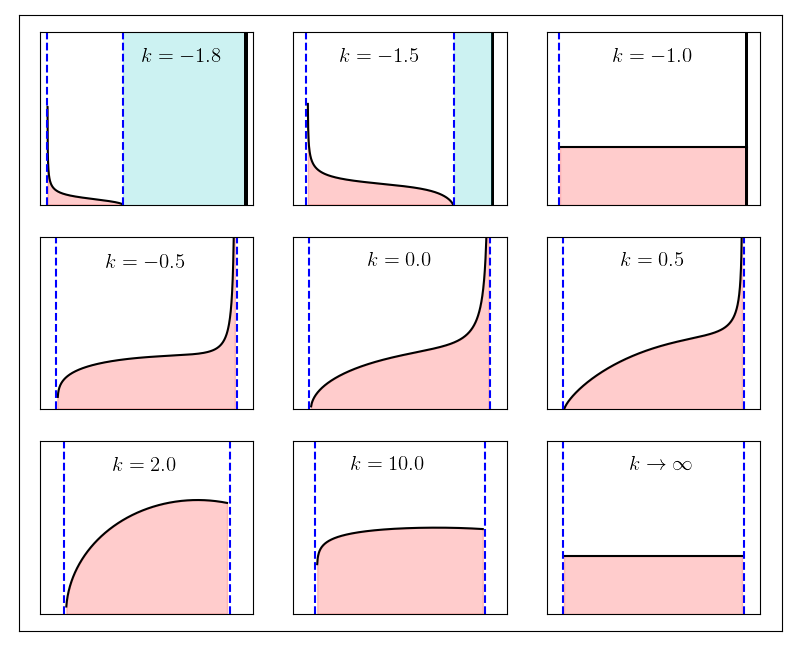}
    \caption{Scaled average density of the constrained gas $\rho_{k}^{\rm{*}}(y)$ vs. $y$ for different values of $k>-2$. The three rows correspond respectively to (i) $k\geq 1$ (third row), (ii) $-1< k < 1$ (second row) and (iii) $-2 < k \leq -1$ (first row). The blue dashed vertical lines indicate the edges of the support. In the second and in the third row, the right edge of the support coincides with the scaled wall position $w$. In the third row the density is a constant at the wall while it diverges in the second row. In both the second and the third row, the density vanishes at the left edge (for $k=10$ in the third row, the true density at the left edge vanishes, though it is not clearly visible due to the compression of the scale). In the first row, the density has an extended bulk part, sandwiched between the two vertical dashed lines and a delta peak at the wall $w$ (shown by a thick solid vertical line). In between, there is hole devoid of particles (shaded cyan region) which disappears for $k=-1$ (the third figure in the first row). 
%
%
    }
    \label{fig:allkwdens}
\end{figure}

Our main result in this paper is the exact computation of the average density of the Riesz gas (characterised by the exponent $k>-2$) in thermal equilibrium in the presence of a wall at position $W$. The effect of the wall is to constrain the particles to stay on the semi-infinite line to the left of the wall. Consider first the unconstrained gas, i.e., without the wall. This is equivalent to placing the wall at $W = + \infty$. 
The scaled density of the unconstrained gas is supported over $[-l_k^{\rm uc},l_k^{\rm uc}]$ (see Fig. \ref{fig:allkdens}) where $l_k^{\rm uc}$ is given in Eq. (\ref{A_1k}). Now imagine bringing the wall from infinity to a finite position $W$. For all $k>-2$, we find that the scaled density of the unconstrained gas remains unchanged as long as the (scaled) position of the wall $w = W/N^{\alpha_k}$ [with $\alpha_k$ given in Eq. (\ref{alpha_k})] is larger than $l_k^{\rm uc}$. In this regime, the particles do not feel the presence of the wall. However, when $w<l_k^{\rm uc}$, the particles feel the presence of the wall and reorganise themselves. This leads to a modification of the mean density and the nature of the modifications depends on the exponent $k$ characterising the range of the interactions. We again find three principal regimes of $k$ (see Fig. \ref{Fig_k}): 
1) $k\geq 1$ (where the interaction is effectively short-ranged), 2) $-1<k<1$ (weakly long-ranged interaction) and 3) $-2 < k \leq -1 $ (strongly long-ranged interaction). The exact form of the modified density in these three regimes are summarised below (see also Fig. \ref{fig:allkwdens}). We obtain these results by employing a saddle-point method in the large-$N$ limit. Finding the analytical solution of this saddle-point equation is the main technical achievement of this paper.

\vspace*{0.5cm}
\noindent{\it Regime 1 ($k \geq 1$): short-ranged interactions}. In this regime the interaction is extremely short-ranged and the effective field-theory becomes local and simple~\cite{agarwal2019harmonically}. Solving the associated saddle-point
equation in the presence of a wall at the scaled position $w=W/N^{\alpha_k}$ with $\alpha_k = k/(k+2)$, we find that, for $w<l_k^{\rm uc}$, the mean density, supported over the finite interval $[-l_k(w),w]$, is given by
\begin{equation}
\rho^*_k(y,w) = 
         A_k \left(l_{k}(w)^2 - y^2\right)^{\frac{1}{k}}\quad, \quad \text{for}  -l_{k}(w)\leq y \leq w, \quad w<l_k^{\rm uc} \;,
\end{equation}
where $A_k$ is given in Eq.~\eqref{A_1k} and the location $-l_{k}(w)$ of the left edge of the support is determined from the normalisation condition $\int_{-l_{k}(w)}^w \rho^*_k(y,w) dy = 1$. This analytical result is verified through Monte-Carlo (MC) simulation in  Fig.~\ref{fig:k1p5a}. The density vanishes at the left edge $-l_k(w)$ while
it approaches a finite value at the right edge, i.e., at the location of the wall (see Fig.~\ref{fig:allkwdens}). In addition, as $w \to -\infty$, the size of the support in scaled units
\begin{eqnarray}\label{def_lengthk}
L_k(w) = w + l_k(w) \;,
\end{eqnarray}
decreases as $L_k(w)\sim|w|^{-\frac{1}{k+2}}$ for a fixed $k\geq1$ (see Fig. \ref{fig:k1p5bc}). 

\vspace*{0.5cm}
\noindent{\it Regime 2 ($-1<k<1$): weakly long-ranged interactions}. In this regime the interaction between two particles at small separation is weaker compared to the previous case ($k>1$), however it is relatively more long-ranged. As a result, the action in the large-$N$ field theory becomes non-local. This modifies the density in a slightly different way compared to the regime 1. We find that the density is still supported on a finite interval $[-l_k(w),w]$ and it vanishes at the left edge $-l_k(w)$. However, at the right edge, i.e., at the wall, the density diverges, though it remains integrable. This is different from the regime 1 where the density approaches a nonzero constant at the wall. We find that the density profile in this regime is explicitly given by 
\begin{align}
    \rho^*_k(y,w) &=
        A_k \frac{(\tilde{l}_k(w)-y)(l_{k}(w)+y)^{\frac{k+1}{2}}}{ (w-y)^{\frac{1-k}{2}}}   \quad, \quad \text{for}  -l_{k}(w)\leq y \leq w \quad \text{,  } \quad w<l_k^{\rm uc}     
        \label{dp_m1<k<1} \\ 
 \text{where}&~~\tilde{l}_k(w) = \frac{1}{2}((k+1) l_{k}(w)+(1-k) w) \;. \label{tilde_l_k>-1}       
\end{align}
Here, $A_k$ is given by Eq.~\eqref{A_1k} and $l_{k}(w)$ is found from normalization of the density. A plot of this expression of the profile is given in Fig.~\ref{fig:k0p5a} where it is also compared with numerical results. As in the case of regime 1, we have studied the support size $L_k(w) = w+\ell_k(w)$ as a function of $w$, shown in Fig.~\ref{fig:k0p5bc}. As $w \to -\infty$, the support size decays algebraically as $L_k(w) \sim |w|^{-\frac{1}{k+2}}$. 

%

\vspace*{0.5cm}
\noindent {\it Regime 3 $(-2<k \leq -1)$: strongly long-ranged interactions.} In this regime, the force between two particles vanishes when two particles get
closer to each other. However, at long separation, the repulsive force increases as a power law $\sim |r|^{|k|-1}$ of the inter-particle distance $|r|$. As a consequence, the associated field-theory is strongly non-local. In the presence of the wall, this leads to a rather exotic scaled density profile consisting of two disjoint pieces separated by a hole: a bulk part, supported over $[-l_k(w),\bar{l}_k(w)]$ and a delta peak with weight $D^*_k(w)$ located at $w > \bar{l}_k(w)$ (see the top row of Fig. \ref{fig:allkwdens}). Thus the hole extends over $[\bar{l}_k(w), w]$ which is devoid of any particle. Moreover, unlike in regimes 1 and 2, where the density vanishes at the left edge $-l_k(w)$, in regime 3, the density actually diverges in an integrable fashion. The presence of the wall is felt over a much wider region in this case, due to the strong non-local nature of the interaction. As the wall is pushed further to the left side, at some critical position $w_c(k)<l_k^{\rm uc}$ the support of the bulk part shrinks to zero and all the particles accumulate at the wall, leading to a single delta function at the wall for $w <w_c(k)$. We find the following explicit expression for the density profile 
\begin{equation}\label{dens-s:klm1}
\rho^*_k(y,w) =     
    \begin{cases}
    A_k \frac{\left(l_{k}(w)+y\right)^{\frac{k+1}{2}} \left(\bar{l}_k(w)-y\right)^{\frac{k+3}{2}}}{(w-y)} 
     \mathbb{I}[-l_{k}(w) < y \leq \bar{l}_k(w)]  &+ \, D^*_k(w)\delta(w-y), \\
 ~~ & \quad \text{for} \quad w>w_c(k) \\
  & \\
     ~~~~~~\delta(w-y), & \quad \text{for} \quad w<w_c(k)
    \end{cases}
\end{equation}
where $\mathbb{I}$ represents the indicator function, $A_k$ is given by Eq.~\eqref{A_1k} and the other constants are given by
\begin{align}
 \bar{l}_k(w) &= \frac{2 w + (k+1) l_{k}(w)}{k+3}, \\
 D^*_k(w) &= \frac{\left(l_{k}(w)-w\right)\left(w+l_{k}(w)\right)^{\frac{k+1}{2}}}{ \left|k\right|\left(k+3\right)} \left(\frac{(k+1)\left(w-l_{k}(w)\right)}{k+3}\right)^{\frac{k+1}{2}},\\
 w_c(k) &= \frac{(k+2) \left| k (k+1)\right| ^{\frac{1}{k+2}}}{k+1}.
\end{align}
The value of $l_{k}(w)$ is again determined from the normalisation condition. The analytical expression in the first line of Eq. (\ref{dens-s:klm1}) is plotted in Fig.~\ref{fig:km1p5} where it is also compared to MC simulations. In this regime 3 of $k$, the support length of the extended part decays as the wall is pushed to the left and goes to zero at a critical wall position $w=w_c(k)$. This can be seen from Fig.~\ref{fig:km1p5bc}. 

Furthermore, in this regime 3 of $k$, we find an interesting first-order phase transition in the density profile as the scaled wall position decreases below a critical value $w^*(k) > w_c(k)$. We find that the actual density profile is a pure delta-function for all $w < w^*(k)$. Thus the solution in the first line of Eq. (\ref{dens-s:klm1}) is actually metastable in the intermediate region $w_c(k)<w<w^*(k)$. This is discussed in Section \ref{derivation2}.

\section{Derivation}
\label{derivation}
For a given configuration of the positions $(x_1,x_2,...,x_N)$ of the particles, we define the empirical density as 
\begin{equation}
    \hat{\rho}_N(x) = \frac{1}{N} \sum_{i=1}^N \delta\left(x-x_i\right) \;. \label{rho_emp}
\end{equation}
We are interested to compute the thermal average of this empirical density for large $N$ which we denote by $\rho_N(x)=\langle \hat{\rho}_N(x)\rangle$. To proceed, we first look at the partition function of the Riesz gas in the presence of wall given by 
\begin{equation}
    Z_k(\beta) = \int_{-\infty}^{W} dx_1  \ldots \int_{-\infty}^{W} dx_N \, e^{-\beta E_k(\{x_i\})}\,  \;.  \label{Z_N(W)} 
\end{equation}
For large $N$ the multiple integrals in the partition function can in principle be done in two steps. First integrate over the microscopic configurations corresponding to a given macroscopic density profile $\rho_N(x)$ and in the second step perform a functional integration over these macroscopic density profiles. After integrating over the microscopic configurations, one obtains, for large $N$, the following functional integral~\cite{agarwal2019harmonically}
\begin{equation}
  Z_k(\beta) \approx \int D[\rho_N] e^{-\beta \mathcal{E}_k\left[\rho_N(x)\right]-  N\int dx \,\, \rho_N(x)\ln\left(\rho_N(x)\right) } \delta \left(\int_{-\infty}^{W} dx \rho_N(x)-1 \right),     \label{Z_k-func}
\end{equation}
where the energy functional $\mathcal{E}_k[\rho_N]$ can be computed from the energy $E_k(\{x_i\})$ given in Eq. (\ref{hamiltonian}). The entropy term in the exponent counts the number of microscopic configurations associated to a macroscopic density $\rho_N(x)$. The delta function ensures that the functional integrals are performed only over normalised density profiles. For large $N$, it has been shown~\cite{agarwal2019harmonically} that the energy functional $\mathcal{E}_k[\rho_N]$ takes the following form, depending on the value of $k$
\begin{equation}\label{mcal(E)}
    \mathcal{E}_k\left[\rho_N(x)\right] \approx \frac{N}{2} \int_{-\infty}^W dx \,\, x^2 \rho_N(x) + 
    \begin{cases}
     \zeta(k) N^{k+1}\int_{-\infty}^W dx \,\, \left[\rho_N(x)\right]^{k+1},& \text{for}\quad k>1\\
      & \\
      N^{2}\ln N\int_{-\infty}^W dx \,\, \left[\rho_N(x)\right]^{2},& \text{for}\quad k=1\\
      & \\
     \frac{\text{sgn}(k)N^2}{2}\int_{-\infty}^W dx' dx \frac{\rho_N(x')\rho_N(x)}{|x-x'|^k} 
      &\text{for}-2<k<1.
    \end{cases}
\end{equation}
%

To determine the scale $L_N$ over which the density varies, we rescale the density as $\rho_N(x)=L_N^{-1}{\rho}_k\left({x}{  L_N^{-1}}\right)$ and substitute this scaling form
in Eq. (\ref{mcal(E)}). The first term corresponding to the confining harmonic potential scales as $N \, L_N^2$. The scaling of the interaction term depends on $k$. For $k >1$, it scales as $N^{k+1}\, L_N^{-k}$, for $k=1$ it scales like $N^2 \, (\ln N)/L_N$ and for $-2<k<1$ it scales as $N^2 L_N^{-k}$\;. Matching the interaction term and the confining term for each $k$, one finds that $L_N \sim N^{\alpha_k}$ where $\alpha_k$ is given in Eq.~(\ref{alpha_k}). In the marginal case $k=1$, $L_N = (N \, \ln N)^{1/3}$. Plugging this scaling ansatz for $\rho_N(x)$ in Eq. (\ref{mcal(E)}), one then gets~\cite{agarwal2019harmonically}
\begin{align}
 \mathcal{E}_k\left[{\rho}_N(x)\right] = \mathcal{B}_N
\tilde{\mathcal{E}}_k\left[{\rho}_k(x L_N^{-1})\right],~\text{where} ~~
\mathcal{B}_N=
\begin{cases}
N^{2 \alpha_k +1}~&\text{for}~k \neq 1 \\ 
N^{5/3} (\ln N)^{2/3}~&\text{for}~k = 1 
\end{cases}
\label{scale-mcal(E)}
 \end{align}
 and the scaled energy functional $ \tilde{\mathcal{E}}_k\left[{\rho}(y)\right]$ takes the following forms
\begin{equation}\label{scaled-field}
    \tilde{\mathcal{E}}_k\left[\rho_k(y)\right] \approx \frac{1}{2} \int_{-\infty}^w dy \,\, y^2 \rho_k(y) +
    \begin{cases}
  \zeta(k) \int_{-\infty}^w dy \,\, \rho_k(y)^{k+1}& \quad \quad \;\; k> 1\\
   \int_{-\infty}^w dy \,\, \rho_k(y)^{2}& \quad \quad \;\; k=1 \\
 \frac{ \text{sgn}(k)}{2} \int_{-\infty}^w \int_{-\infty}^w dy' dy \,\, \frac{\rho_k(y)\rho_k(y')}{|y'-y|^k} & -2<k<1 \,,
    \end{cases}
\end{equation}
with $w=W/L_N$. Substituting (\ref{scale-mcal(E)}) in the expression (\ref{Z_k-func}) for the partition function, one finds that the energy scale
$\mathcal{B}_N$ is much bigger than the scale of the entropy, since $\mathcal{B}_N \gg N$ for large $N$ and fixed $\beta$. Hence, neglecting the entropy term and using the integral representation of the delta function $\delta(x) = \int_\Gamma \frac{d\mu}{2\pi i}e^{\mu x}$ where $\Gamma$ runs along the imaginary axis in the complex $\mu$-plane, we rewrite the  partition function in Eq.~\eqref{Z_k-func} as 
\begin{equation}\label{partition-field}
    Z_k(\beta) = \int d\mu \int \mathcal{D}[\rho_k]~\text{exp}\left[-\beta  \mathcal{B}_N \Sigma_k\left[ \rho_k(y),\mu\right] + o(\mathcal{B}_N)\right] ,
\end{equation}
where  $o(\mathcal{B}_N)$ represents terms of order smaller that $\mathcal{B}_N$ (subdominant) and the action $\Sigma_k\left[\rho_k(y),\mu\right]$ is given by 
\begin{equation}\label{action}
    \Sigma_k\left[\rho_k(y),\mu\right] = \left(\tilde{\mathcal{E}}_k\left[\rho_k(y)\right] - \mu \left(\int dy \rho_k(y) -1\right)\right) \;,
\end{equation}
with $\tilde{\mathcal{E}}_k\left[\rho_k(y)\right]$ given in Eq. (\ref{scaled-field}).  

The integrals in Eq.~\eqref{partition-field} can be performed using saddle point method in which one requires to minimise the action $\Sigma_k[{{\rho}}_k(y), \mu]$ in Eq. (\ref{action}) to find the saddle point density $\rho^*_k(y,w)$ and the chemical potential $\mu_k^*$. The saddle point equations read
\begin{align}
    \frac{\delta \Sigma_k\left[\rho_k\left(y\right), \mu \right]}{\delta \rho_k\left(y\right)}\Bigg|_{\substack{\rho_k\left(y\right) = \rho^*_k\left(y, w\right)\\\mu=\mu_k^*}} &= 0 \label{saddle-den_eqn}\\
    \frac{\partial \Sigma_k\left[\rho_k\left(y\right), \mu \right]}{\partial \mu}\Bigg|_{\substack{\rho_k\left(y\right) = \rho^*_k\left(y, w\right)\\\mu=\mu_k^*}} &= 0 \;. \label{saddle-mu_eqn}
\end{align}
Note that the second equation above is equivalent to the normalization condition $\int dy \rho_k(y) = 1$. In the limit $N \to \infty$, the saddle point density clearly coincides with the average density.

\subsection{Regime 1 ($k \geq 1$): short-ranged interactions}
In this regime the interaction energy falls so quickly with increasing separation that it  effectively acts as short-ranged and consequently the energy functional $   \tilde{\mathcal{E}}_k\left[\rho_k(y)\right] $ becomes local in the leading order for large $N$ (see Eq.~\eqref{scaled-field}). The saddle point equation~\eqref{saddle-den_eqn} becomes 
\begin{equation}\label{kg1-chem-eq}
 \mu_k^* = \frac{y^2}{2}  + (k+1) \zeta(k)\left[\rho^*_k \left(y, w \right) \right]^{k},
\end{equation}
for $k>1$ (the case $k=1$ is treated separately below). This equation of course is valid for $y$ belonging to the support of the density. To determine the support, we solve 
Eq.~(\ref{kg1-chem-eq}) explicitly, giving
\begin{equation}\label{kg1dens}
    \rho^*_k(y, w) = A_k \left(2 \mu_k^* - y^2\right)^{\frac{1}{k}}  \;,
\end{equation}
where $A_k$ is given in the first line of Eq.~\eqref{A_1k}. This density is real and nonzero for $-\sqrt{2 \mu_k^*} < y < + \sqrt{2 \mu_k^*}$. Now there are two possible situations: (i) when $w>\sqrt{2 \mu_k}$ and (ii) when $w<\sqrt{2 \mu_k}$. Consider the situation (i) first. In this case, the density is given by Eq. (\ref{kg1dens}) and
is supported over $[-\sqrt{2 \mu_k^*}, + \sqrt{2 \mu_k^*}]$. The only unknown is $\mu_k^*$ which is fixed by the normalization condition $\int_{-\sqrt{2 \mu_k^*}}^{\sqrt{2 \mu^*_k}}  \rho^*_k(y, w)\, dy = 1$. It is easy to show that it gives $\sqrt{2 \mu_k^*} = l_k^{\rm uc}$ where $l_k^{\rm uc}$ is given in Eq. (\ref{A_1k}). In this case, the density in (\ref{kg1dens}) is precisely the unconstrained density given in Eq. (\ref{rho_uc}). Thus we conclude that for $w > l_k^{\rm uc}$ the unconstrained density is not affected by the presence of the wall. 

\begin{figure}[t]
    \centering
    \includegraphics[width=0.8\linewidth]{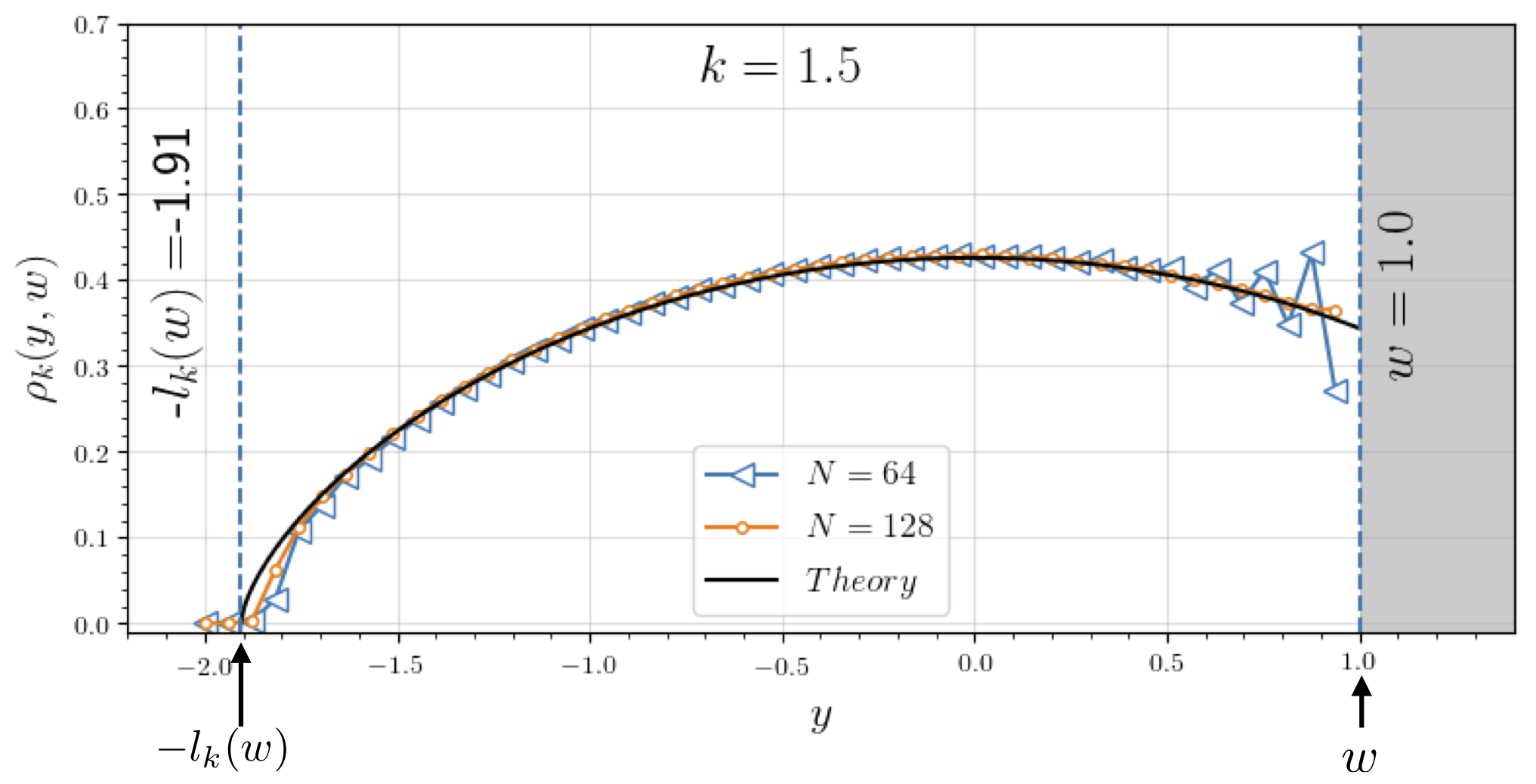}
    \caption{\textit{Regime $1$} ($k\geq 1$): Comparison between MC simulations (symbols) and the theoretical expression given in Eq.~\eqref{kg1dens} (solid line) of the scaled density profile of the Riesz gas in the presence of the hard wall at $w=1.0$ for $k=1.5$. The parameters used in this plot are $J=1$ and $T=1$. The dashed vertical line on the left marks the left edge $-l_k(w)$ (shown by an arrow) of the density while the wall is located at $w$ (shown also by an arrow). 
    }\label{fig:k1p5a}
\end{figure}

We next consider the case (ii) above, i.e., when $w < \sqrt{2 \mu_k^*}$. In this case, the support of the density in Eq. (\ref{kg1dens}) is over $[-\sqrt{2 \mu_k^*},w]$. Thus, unlike in case (i) above, the density does not vanish at the upper edge $w$ of the support (see Fig. \ref{fig:k1p5a}) and it reads
\begin{equation}\label{kg1densii}
    \rho^*_k(y, w) = A_k \left(2 \mu_k^* - y^2\right)^{\frac{1}{k}}  \;,\; -\sqrt{2 \mu_k^*} \leq y \leq w \;. 
\end{equation}
Setting $\sqrt{2\mu_k^*} = l_k(w)$, the density is then supported over $[-l_k(w),w]$. The only unknown $l_k(w)$ is then fixed by the normalization condition $\int_{-l_k(w)}^{w}  \rho^*_k(y, w)\, dy = 1$. Substituting the density from Eq. (\ref{kg1densii}), the normalization condition can be expressed in terms of an auxiliary variable 
$m_k = \frac{w+l_k(w)}{2 l_{k}(w)}$
\begin{align}
(2m_k-1) \left(\frac{B(\gamma_k +1, \gamma_k+1)}{B(m_k; \gamma_k +1, \gamma_k+1)}\right)^{\alpha_k}=\frac{w}{l_k^{\rm uc}}
  \label{l_{k}(w)-k>1}
\end{align}
where $B(m_k; a, b) = \int_0^{m_k} u^{a-1}(1-u)^{b-1}\, du$ is the incomplete Beta function and we recall that $\gamma_k = 1/k$ and $\alpha_k = k/(k+2)$. The variable $m_k$ lies in the range $[0,1]$. Solving Eq. (\ref{l_{k}(w)-k>1}) gives $m_k$, which in turn fixes the unknown constant $l_k(w)$. Let us investigate two limiting cases. First consider the limit $w \to l_k^{\rm uc}$ from the left. In this case the right hand side of Eq. (\ref{l_{k}(w)-k>1}) approaches $1$ and therefore $m_k \to 1$ in this limit, i.e., $l_k(w) \to l_k^{\rm uc}$ as expected. In the opposite limit where $w \to - l_k(w)$ (i.e., in the limit of vanishing support size $L_k(w)= w+ l_k(w)\to 0$, which happens when $w \to -\infty$), the variable $m_k \to 0$. Using the small $m_k$ behavior of $B(m_k; a, b) \sim m_k^a$ in Eq. (\ref{l_{k}(w)-k>1}), it is easy to verify that the support length $L_k(w) = w + l_k(w) \sim |w|^{-1/(k+1)}$, as $w \to - \infty$.

\begin{figure}[t]    
\centering
    \includegraphics[width=0.8\linewidth]{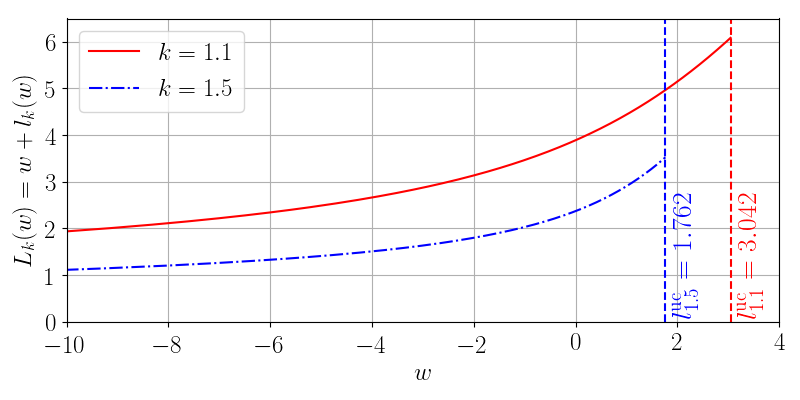}
    \caption{\textit{Regime $1$} ($k \geq 1$): Plot of the support size $L_k(w)=w+l_k(w) = \frac{2m_k w}{2m_k-1}$ as a function of $w\leq l_k^{\rm uc}$, obtained by solving Eq.~\eqref{l_{k}(w)-k>1} for $m_k$. $L_k(w)$ decreases monotonically with decreasing $w$, shown for two different values of $k$. As $w \to - \infty$, the support size decreases algebraically as  $L_k(w) \sim |w|^{-1/(k+1)}$. The vertical dashed lines indicate the positions of the unconstrained right edge $l_k^{\rm uc}$ for the two values of $k$.} 
    \label{fig:k1p5bc}
\end{figure}

The density profile of the constrained gas in Eq. (\ref{kg1densii}) is plotted (solid line) in  Fig.~\ref{fig:k1p5a} where it is compared with numerical simulations (symbols) for different values of $N$. We observe that for increasing $N$ the numerical density profile converges to the analytical expression.
Clearly, the density is nonzero at the wall while it vanishes at the left edge as $\sim (l_{k}(w)-|y|)^{\gamma_k}$ with $\gamma_k=1/k$, as in the unconstrained case. As $w$ decreases below $l_{k}^{\rm uc}$, the gas is pushed to the left and, as argued above, the support size $L_k(w) = w + l_k(w)$ shrinks algebraically $L_k(w)\sim |w|^{-\frac{1}{1+k}}$ as $w \to - \infty$. This result is verified in simulations in Fig.~\ref{fig:k1p5bc} where we plot the support length $L_k(w)$ as a function of $w$ for given $k$. 
Following the same procedure for $k=1$, we find that the density profile is  given by the same form as for $k>1$
\begin{equation}\label{k1dens}
    \rho^*_1(y, w) = A_1\, \left(l_1(w)^2 - y^2\right), \quad \text{for} \quad -l_1(w) \leq y \leq w, \quad w<l_1^{\rm uc} \;,
\end{equation}
with the prefactor $A_1 = 1/4$ [see the second line of Eq. (\ref{A_1k})] and $l_1(w)$ is determined for the normalization condition.

\subsection{Regime 2: Weakly long-ranged interactions $(-1 < k < 1)$}
\label{derivation1}

In this regime of $k$, the interaction forces decay slower with increasing inter-particle separation compared to the previous short-ranged regime. The energy functional in this regime is given by the third line in Eq.~\eqref{scaled-field}. Using this in the saddle point equation~\eqref{saddle-den_eqn} we get 
\begin{equation}\label{kl1 chem eq}
\mu_k^* = \frac{y^2}{2}  + \text{sgn}(k)\int_{-\infty}^{w} dy' \,\, \frac{\rho^*_k\left(y', w\right)}{|y'-y|^k}.
\end{equation}
\noindent
To solve this equation, we note that the first term on the right hand side grows arbitrarily for large negative $y$ whereas the second term can at maximum grow as $y^{1+|k|}$ (for $k<0$). Since $\mu_k^*$ is a constant the Eq.~\eqref{kl1 chem eq} can be valid only for density profiles with  finite support $(-l_{k}(w), w)$.
Taking a derivative with respect to $y$ on both sides of Eq.~\eqref{kl1 chem eq} we get
\begin{equation}\label{kl1 int eq}
P.V.\int_{-l_{k}(w)}^{w} \frac{\rho^*_k(y', w) \text{sgn}(x-y)}{\left| y-y'\right| ^{k+1}} \, dy'=-\frac{y}{\left| k\right|} \quad , \quad -l_k(w)\leq y \leq w \;,
\end{equation}
where $P.V.$ represents the principal value. Note that this integral is interpreted in principal value sense only for $0\leq k <1$, but for $-2<k<0$ it is considered as normal integral. We need to solve the integral equation~\eqref{kl1 int eq} to obtain the desired density. We can simplify the calculations by shifting the coordinates to the left edge and scaling with the length of the support $L_k(w) = w+l_{k}(w)$, {i.e.}, by making the transformation $z=\frac{y+l_{k}(w)}{L_k(w)}$. Since the density is normalised it is expected to take the scaling form
\begin{equation}
\rho^*_k(y, w)= \frac{1}{L_k(w)} \phi_k\left(\frac{y+l_{k}(w)}{L_k(w)},w\right) \;,
\end{equation}
where $\phi(z,w)$ is now supported over $z \in [0,1]$ and satisfies the Sonin equation 
\begin{equation}\label{kl1 int eq 1}
P.V. \int_{0}^{1} \frac{\phi_k(z', w)\text{sgn}(z'-z)}{\left| z-z'\right| ^{k+1}} \, dz'=h_k(z) \quad, \quad 0 \leq z \leq 1 \;,
\end{equation}
with $h_k(z)= \mathcal{A}_k\left(z-q_k(w)\right)$, $\mathcal{A}_k =  -\frac{[L_k(w)]^{k+2}}{\left| k\right| }$ and $q_k(w) = \frac{l_k(w)}{L_k(w)}$. The only unknown so far is $l_k(w)$. 
\begin{figure}[t]
    \centering
    \includegraphics[width=0.8\linewidth]{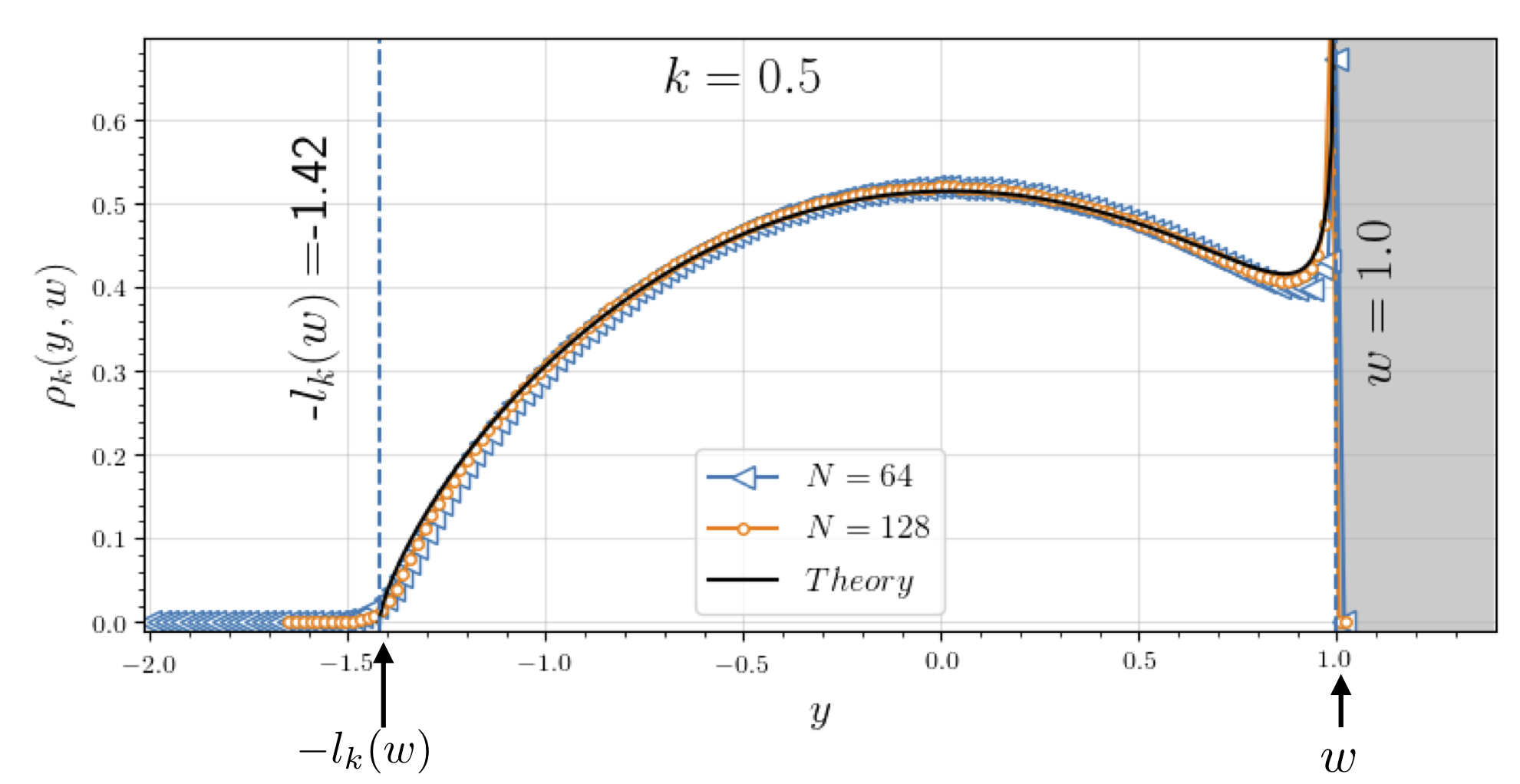}
    \caption{\textit{Regime $2$} ($-1<k<1$): Comparison between MC simulations (symbols) and the theoretical expression given in Eq.~\eqref{kl1dens} (solid line) of the scaled density profile of the Riesz gas in the presence of the hard wall at $w=1.0$ for $k=0.5$. The parameters used in this plot are $J=1$ and $T=1$. The dashed vertical line on the left marks the left edge $-l_k(w)$ (shown by an arrow) of the density while the wall is located at $w$ (shown also by an arrow). The density diverges at the wall as $\sim (w-y)^{(k-1)/2}$. 
    }
     \label{fig:k0p5a}
\end{figure}
Fortunately,  the Sonin equation can be inverted for arbitrary source function $h_k(z)$ and the general solution is given by~\cite{buldyrev2001properties}
\begin{align}\label{sol:sonin}
    \phi_k(z,w) &= C_0\big(z(1-z)\big)^{\frac{k-1}{2}} + u_k(z),~~\text{with} 
\end{align}
\begin{equation}\label{sol:sonin1}
u_k(z)=\frac{2 A_k |k|}{B\left(\frac{k+1}{2},\frac{k+1}{2}\right)}z^{\frac{k-1}{2}} \frac{\partial }{\partial z} \left(\int_z^{1} t^{-k} (t-z)^{\frac{k+1}{2}} \frac{\partial }{\partial t }\int_0^t h(y)y^{\frac{k+1}{2}} (t-y)^{\frac{k-1}{2}} \, dy \, dt\right),
\end{equation}
where $C_0$ is an arbitrary constant and $A_k$ is given in Eq. (\ref{A_1k}). In Eq. (\ref{sol:sonin}), the first term represents the general solution of the homogenous part of the equation (with $h_k(z)=0$), while the second term $u_k(z)$, given explicitly in Eq. (\ref{sol:sonin1}), represents a particular solution of the full inhomogeneous equation (\ref{kl1 int eq 1}). Inserting the explicit form of $h_k(z)$ [given after Eq.~\eqref{kl1 int eq 1}] in Eq.~\eqref{sol:sonin1} and performing the integral (detailed in~\ref{apndkl1}), we obtain the full general solution
\begin{equation}\label{phikl11}
\begin{split}
    \phi_k(z, w) = \big(z(1-z)\big)^{\gamma_k-1}\Big[C_0-A_k |k| \mathcal{A}_k (1-z)\big(z- \gamma_k(2q_k(w)-1)\big)\Big] \;, 
\end{split}
\end{equation}
where $\gamma_k = (k+1)/2$. The only unknown parameters so far are $C_0$ and $l_k(w)$.

\begin{figure}[t]    
\centering
        \includegraphics[width=0.8\linewidth]{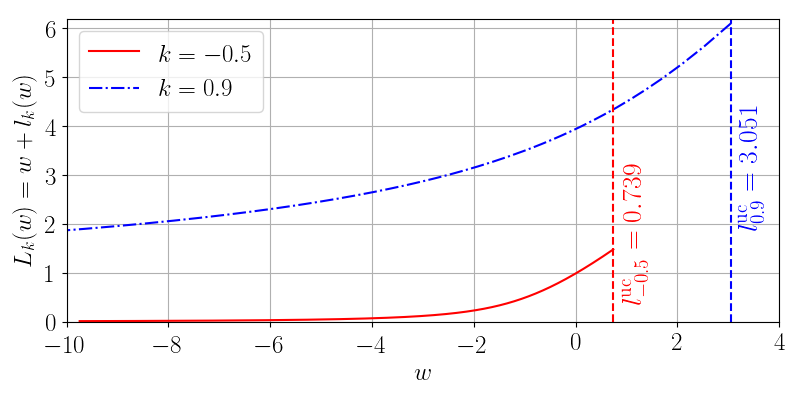}
    \caption{\textit{Regime $2$} ($-1<k<1$): Plot of the support size $L_k(w)=w+l_k(w) = \frac{2 \gamma_k\, w}{1+\gamma_k-g_k(w)}$ as a function of $w\leq l_k^{\rm uc}$,  obtained by solving Eq.~(\ref{l_{k}(w)-k<1}) for $g_k(w)$. $L_k(w)$ decreases monotonically with decreasing $w$, shown for two different values of $k$. As $w \to - \infty$, the support size decreases algebraically as  $L_k(w) \sim |w|^{-1/(k+1)}$. The vertical dashed lines indicate the positions of the unconstrained right edge $l_k^{\rm uc}$ for the two values of $k$.}
    \label{fig:k0p5bc}
\end{figure}

As in the regime 1, there are two possible scenarios, depending on the value of $w$ compared to the unconstrained right edge $l_k^{\rm uc}$. If $w > l_k^{\rm uc}$, it turns out that the constant $C_0=0$ and the density is the same as the unconstrained density supported over $[-l_k^{\rm uc}, +l_k^{\rm uc}]$ and is given in Eq. (\ref{rho_uc}). The situation however is drastically different for $w < l_k^{\rm uc}$. In this case, it turns out that the constant $C_0$ is nonzero and is determined as follows.   
To fix $C_0$ in this case, we need to use some informations about the density profile from numerical simulations (see Fig. \ref{fig:k0p5a}). From the simulations, we see that the density vanishes at the left edge (corresponding to $z=0$ in the shifted coordinate), while it diverges at the right edge at $w$ (corresponding to $z=1$ in the shifted coordinate). From Eq. (\ref{phikl11}), since $\gamma_k - 1 = (k-1)/2 < 0$ here for $k<1$, it follows that if the density has to vanish at $z=0$, the term in the square bracket must vanish at $z=0$. This fixes the constant $C_0 = -A_k |k| \mathcal{A}_k(\gamma_k-1)(2q_k(w)-1)$. Hence $\phi_k(z,w)$ becomes  
\begin{equation}
    \phi_k(z,w) =A_k |k| \mathcal{A}_k z^{\gamma_k} (1-z)^{\gamma_k-1} (g_k(w)-z)\mathbb{I}[0<z<1]
\end{equation}
where 
\begin{eqnarray} \label{def_gk}
g_k(w) = \gamma_k\left(2q_k(w)-1\right)+1 = \frac{(k+3)l_k(w)+(1-k)w}{2(w+l_k(w))}\;,
\end{eqnarray}
and we used $q_k(w) = l_k(w)/(w+l_k(w))$. Note that $g_k(w)$ in Eq. (\ref{def_gk}) can also be expressed in
terms of the support length $L_k(w) = w+l_k(w)$ as
\begin{eqnarray} \label{def_gk2}
g_k(w) = \frac{k+3}{2}- \frac{(k+1)\,w}{L_k(w)} \;.
\end{eqnarray}
Thus, finally, the density in terms of the original coordinate $y$ reads
\begin{equation}\label{kl1dens}
{\rho}^* _k(y, w) =A_k (l_{k}(w)+y)^{\frac{k+1}{2}} (w-y)^{\frac{k-1}{2}} \left(\tilde{l}_k(w)-y \right), ~~ \text{for} ~~ -l_{k}(w) \leq y \leq w\;,\; ~~ w<l_k^{\rm uc}
\end{equation}
where $\tilde{l}_k(w) = \frac{1}{2}((k+1) l_{k}(w)+(1-k) w)$ as given in Eq.~\eqref{tilde_l_k>-1}. The only remaining unknown $l_k(w)$ 
is then determined from the normalization condition $\int_{-l_{k}(w)}^w \rho^*_k(y, w)\, dy =1$. The normalization condition can be conveniently expressed
in terms of $g_k(w)$ defined in Eq. (\ref{def_gk}) as
\begin{align}
 \label{l_{k}(w)-k<1}
 \left(\frac{\gamma_k+1-g_k(w)}{\gamma_k}\right)\left(g_k(w)\left(2+\frac{1}{\gamma_k}\right)-\left(1+\frac{1}{\gamma_k}\right)\right)^{-\alpha_k} = \frac{w}{l_k^{\rm uc}} \;,
\end{align}
where we recall that $\alpha_k = 1/(k+2)$ and $\gamma_k = (k+1)/2$. This equation is the analogue of Eq. (\ref{l_{k}(w)-k>1}) in the regime 1. For a given $w$ and $k$, we solve this equation numerically to get $g_{k}(w)$ which, via Eq. (\ref{def_gk}), in turn fixes the only remaining unknown constant $l_k(w)$. Once $l_k(w)$ is fixed, Eq. (\ref{kl1dens}) then provides the exact density profile of the constrained gas. We verify that in the limit $k \to 0$ our results recover the known results for the Dyson's log gas in the presence of a wall \cite{dean2006large, dean2008extreme}. The numerical results from MC simulation for the density profile are in perfect agreement with our analytical predictions, as shown in Fig.~\ref{fig:k0p5a} for a representative value $k=-0.5$ in this regime.

One can show that the solution for $g_k(w)$ in Eq. (\ref{l_{k}(w)-k<1}) lies in the range $[1,+\infty)$. It is easy to see that when $w$ approaches $l_k^{\rm uc}$, the right hand side of Eq. (\ref{l_{k}(w)-k<1}) approaches $1$ and consequently $g_k(w) \to 1$. From Eq. (\ref{def_gk}), we see that $l_k(w) \to w$ and since $w \to l_k^{\rm uc}$ it follows 
that $l_k(w) \to l_k^{\rm uc}$, i.e., to the right edge of the unconstrained gas. In contrast, when $w \to -\infty$, it is easy to check from Eq. (\ref{l_{k}(w)-k<1}) that $g_k(w)$ diverges as $g_k(w) \sim |w|^{\frac{k+2}{k+1}}$. Substituting this behavior in Eq. (\ref{def_gk2}), we see that $L_k(w) \sim |w|^{-\frac{1}{k+1}}$ as $w \to - \infty$. In Fig. \ref{fig:k0p5bc}, we plot the support size $L_k(w)$ as a function of $w$
for $k=-0.5$ and $k=0.9$. 

We see from Eq. (\ref{kl1dens}) that the density at the wall diverges as $\rho_k^*(y,w)\sim (w-y)^{(k-1)/2}$ (since $k<1$). Thus the divergence becomes stronger as $k$ decreases and at $k=-1$ it becomes non-integrable, signalling a breakdown of the validity of the solution in Eq. (\ref{kl1dens}). This calls for a different analysis for $-2<k\leq -1$, which we carry out in the next subsection.

\subsection{Regime 3: Strongly long-ranged interactions ($-2 < k \leq -1$)}
\label{derivation2}
\begin{figure}[t]
    \centering
    \includegraphics[width=0.8\linewidth]{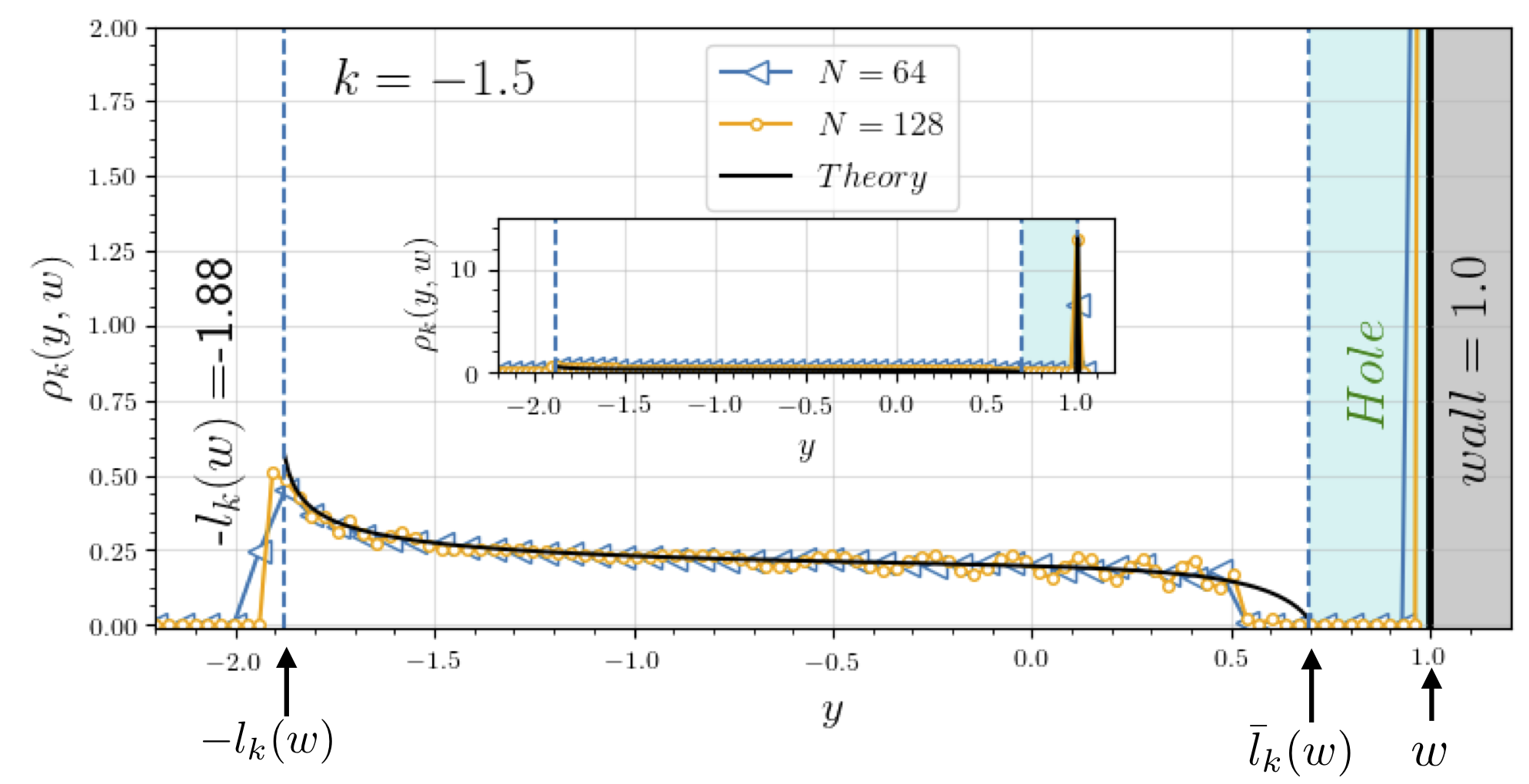}
    \caption{\textit{Regime $3$} ($-2<k \leq -1$): Scaled density profile for $w=1$ and $k=-1.5$: comparison between MC simulations (symbols) and the theoretical expression given in Eq.~\eqref{dens:ansatz} with $\rho_b(y,w) = \rho_b^*(y,w)$ in Eq. (\ref{dens:klm1}) and $D_k(w) = D_k^*(w)$ in Eq. (\ref{dwkm1p5}) (solid line). The parameters used in this plot are $J=1$ and $T=1$. The dashed vertical line on the left (right) marks the left (right) edge $-l_k(w)$ ($\bar{l}_k(w)$) shown by an arrow. The cyan shaded area is the hole region and the thick black line at $y=w=1.0$ is the delta function.}
    \label{fig:km1p5}
\end{figure}

In this regime not only the interaction energy but also the interaction force is zero at vanishingly small separation. 
As a consequence of this the density in this regime, in the presence of a wall, displays interesting features as seen in Fig.~\ref{fig:km1p5} where we plot constrained density profile $\rho^*_k(y, w)$ as a function of $y$ obtained from numerical simulations. Interestingly, in this case, for $k<-1$, the average density profile, supported over a finite range, consists of two disjoint parts with a region devoid of the particles (hole) in between them. One part corresponds to a very high density (a delta peak) at the position of the wall (see the inset in Fig.~\ref{fig:km1p5}). The other part has an extended profile which vanishes at the right edge bordering the hole and has an integrable divergence at the left edge (see Fig. \ref{fig:km1p5}). Strictly for $k=-1$ ($1d$OCP) the hole disappears and the extended bulk merges with the delta-peak~\cite{dhar2018extreme}. These observations suggest an ansatz for the scaled density profile of the form
\begin{equation}\label{dens:ansatz}
\rho_k(y, w)=\rho_b(y, w)\mathbb{I}[-l_k(w)\leq y \leq \bar{l}_k(w)]+ D_k(w) \delta (w-y) \;,
\end{equation}
where $-l_k(w)<\bar{l}_k(w)<w$. The extended part of the density $\rho_b(y, w)$ (where the subscript $b$ refers to the bulk density) is assumed to be supported over $(-l_k(w), \bar{l}_k(w))$. Here $\bar l_k(w)$ denotes the right edge of the extended density profile, or equivalently the left edge of the hole. Hence the
hole is over the region $y\in (\bar{l}_k(w), w)$. The amplitude $D_k(w)$ of the delta-function just denotes the fraction of particles in the delta-peak. In fact, with this ansatz (\ref{dens:ansatz}) the normalisation condition reads
\begin{eqnarray} \label{norm_cond}
\int_{-l_k(w)}^{\bar{l}_k(w)} \rho_b(y, w) \, dy + D_k(w) = 1 \;.
\end{eqnarray}
The next step is to substitute this ansatz (\ref{dens:ansatz}) in the expression for the scaled energy function in the third line of Eq. (\ref{scaled-field}). It reads
\begin{eqnarray} \label{ener_func_3}
&&\hspace*{-2.cm}\tilde{\mathcal{E}}_k\left[\rho_k(y)\right] \approx \frac{1}{2} \int_{-l_k(w)}^{\bar{l}_k(w)} dy \,\, y^2 \rho_b(y, w) -\frac{1}{2} \int_{-l_k(w)}^{\bar{l}_k(w)} \int_{-l_k(w)}^{\bar{l}_k(w)} dy' dy \,\, \frac{\rho_b(y,w)\rho_b(y',w)}{|y'-y|^k} \nonumber \\
&& + D_k(w) \left[ \frac{w^2}{2} -\int_{-l_k(w)}^{\bar{l}_k(w)} \frac{\rho_b(y,w)}{|w-y|^{k}}\; dy  \right] \;.
\end{eqnarray}
The first two terms represent the energy of the particles in the extended part with density $\rho_b(y,w)$. The third term represents the energy of the particles localised in the
delta-function -- it has two parts: the first part $D_k(w)\, w^2/2$ represents the potential energy of these particles while the second part represents the long-ranged interaction energy between these particles and the extended bulk with density $\rho_b(y,w)$, separated by the hole. Note that the interaction energy between the particles localised at $w$ does not contribute as it vanishes identically for $k<0$, which is the case in this regime 3.

The goal is now to minimise this scaled energy functional in (\ref{ener_func_3}) by varying $\rho_b(y,w)$. Note that the amplitude of the delta-peak $D_k(w)$ is automatically fixed by the normalization condition (\ref{norm_cond}). Hence the optimisation will be only with respect to $\rho_b(y,w)$, and not $D_k(w)$ independently. Taking a functional derivative with respect to $\rho_b(y,w)$ subject to the constraint (\ref{norm_cond}) enforced by a Lagrange multiplier $\mu_k$, we get    
\begin{align}
 \mu_k^*&=\frac{y^2}{2}- \int_{-l_{k}(w)}^{\bar{l}_k(w)} \frac{\rho^*_b(y', w)}{\left| y-y'\right| ^k} \, dy'- D^*_k(w) (w-y)^{-k} \;,\label{chem-pote}
 \end{align}
where the subscript `$*$' indicates the optimal value of the parameters and the density. The optimal density $\rho^*_b(y,w)$ has thus two unknown parameters $-l_k(w)$ and $\bar{l}_k(w)$ and we recall that the constant $D_k(w)$ is fixed from the normalization condition (\ref{norm_cond}).

To proceed further we take a derivative of Eq.~\eqref{chem-pote} with respect to $y$ and get
\begin{equation}\label{longklm1:integraleqn}
\int_{-l_{k}(w)}^{\bar{l}_k(w)} \frac{\rho^*_b(y', w)\text{sgn}(y'-y)}{\left| y-y'\right| ^{k+1}} \, dy'=-\frac{y}{\left| k\right| }-D^*_k(w) (w-y)^{-(k+1)}.
\end{equation}
This can be simplified after a change of variable $z = \frac{y+l_k(w)}{\tilde{L}_k(w)}$ where $\tilde{L}_k(w) = \bar{l}_k(w)+l_k(w)$ is the size of the support. In this shifted and scaled coordinate the density takes the scaling form 
\begin{equation}
\rho^*_b(y, w)= \frac{1}{\tilde{L}_k(w)} \phi_k\left(\frac{y+l_{k}(w)}{\tilde{L}_k(w)},w\right) \;,
\label{scalingdensklm1}
\end{equation}
where $\phi_k(z, w)$ satisfies the following equation
\begin{equation}
    \int_0^1 dz' \frac{\text{sgn}(z'-z)}{|z'-z|^{k+1}} \phi_k(z', w) = h_k(z) \quad, \quad 0 \leq z \leq 1 \;,
    \label{longklm1:integraleqn2}
\end{equation}
with $h_k(z)= \mathcal{A}_k(z-q_k) + \mathcal{B}_k(g_k-z)^{-(k+1)}$. The constants are 
\begin{eqnarray} \label{constants_3}
\hspace*{-2cm}\mathcal{A}_k = - \frac{\left[\tilde{L}_k(w)\right]^{k+2}}{|k|} \;, \;g_k(w) = \frac{w+l_k(w)}{\tilde{L}_k} \;, \; q_k(w) = \frac{l_k(w)}{\tilde{L}_k(w)} \;, \; \mathcal{B}_k = -D^*_k(w) \;,
\end{eqnarray} 
where we recall that $\tilde L_k(w) = \bar{l}_k(w) + l_k(w)$. This equation (\ref{longklm1:integraleqn2}) looks similar to (\ref{kl1 int eq 1}) in regime 2. However, there is no principal value ($P.V.$) in Eq. (\ref{longklm1:integraleqn2}). This is due to the fact that for $k \leq -1$ the integrand is not singular inside the support. 

The integral equation~\eqref{longklm1:integraleqn2} can be solved exactly using Sonin inversion formula given in Eq.~\eqref{sol:sonin}. After a long calculation presented in~\ref{apndklm1}, we find
\begin{equation}
\begin{split}
    \phi_k(z, w) = \big(z(1-z)\big)^{\gamma_k-1}\Bigg[ C_0& - A_k |k| (1-z) \mathcal{A}_k \left(\gamma_k \big(1-2q_k(w)\big)+z\right)\\& - A_k |k| \frac{1-z}{g_k(w)-z} \mathcal{B}_k \frac{2\gamma_k g_k(w)}{(g_k(w)(g_k(w)-1))^{\gamma_k}}
     \Bigg],
\end{split}
\label{phi:klm1}
\end{equation}
where $C_0$ is an arbitrary constant and $\gamma_k = (k+1)/2$. Thus so far, we have three unknown constants characterising the optimal density: $l_k(w)$, $\bar{l}_k(w)$ and $C_0$. To fix these three unknowns we proceed as follows. 

We start by fixing $C_0$. As $z \to 1$ in Eq. (\ref{phi:klm1}), the density $\phi_k(z,w) \sim C_0(1-z)^{(k-1)/2}$. Since $k\leq -1$ we see that the density has a non-integrable divergence at $z=1$ unless $C_0=0$. Since the density is normalizable, $C_0=0$ is the only possible choice. Setting $C_0=0$ in Eq. (\ref{phi:klm1}), we get
\begin{eqnarray}\label{phi_k1}
&&\phi_k(z,w) = -A_k |k| \mathcal{A}_k\frac{z^{\gamma_k-1}\left(1-z\right)^{\gamma_k}}{g_k(w)-z} \nonumber \\
&& \times \left[\left(\gamma_k \big(1-2q_k(w)\big)+z\right)(g_k(w)-z) + \frac{\mathcal{B}_k}{\mathcal{A}_k} \frac{2\gamma_k g_k(w)}{(g_k(w)(g_k(w)-1))^{\gamma_k}}\right] \;,
\end{eqnarray}
where we recall again that $\gamma_k = (k+1)/2 < 0$ in this regime 3. Let us first look at the edge at $z=1$. The term $(1-z)^{\gamma_k}$ clearly diverges at the right edge, where $z \to 1$. On the other hand, from MC simulations, we see the density always vanishes at this edge. This means that the term inside the square bracket in the second line of Eq. (\ref{phi_k1}) must vanish as $z \to 1$. Secondly, investigating the $z\to 0$ limit in Eq. (\ref{phi_k1}), we see that the amplitude diverges as $z^{(k-1)/2}$ which leads to a non-integrable divergence. Hence the term inside the square bracket in the second line of Eq. (\ref{phi_k1}) must also vanish as $z \to 0$. Note that the square bracket on the second line of Eq. (\ref{phi_k1}) is a polynomial in $z$ of degree $2$, and hence it must of the form $z(1-z)$ in order to satisfy the behavior at both edges $z=0$ and $z=1$. 
%
\begin{figure}[t] 
\centering
    \includegraphics[width=0.8\linewidth]{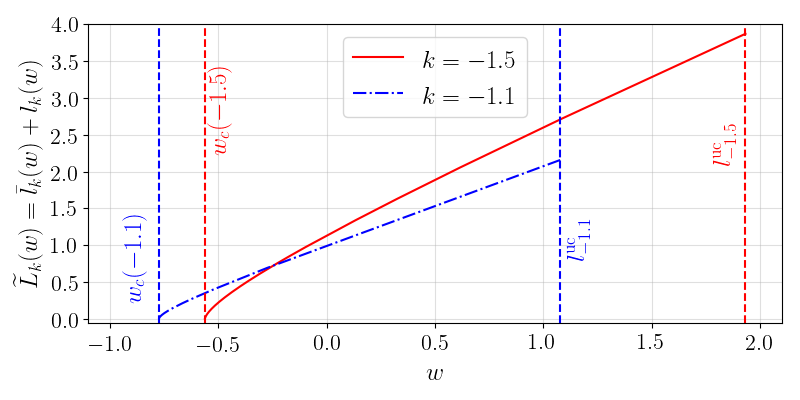}
    \caption{\textit{Regime $3$} ($-2<k\leq -1$): Plot of the support size $\tilde L_k(w)=\bar{l}_k(w)+l_k(w)$ in Eq. (\ref{Lktil}) (with $g_k(w)$ determined from Eq. (\ref{l_{k}(w)-k<-1})). $\tilde L_k(w)$ decreases monotonically with decreasing $w$ and vanishes at $w=w_c(k)$ given in Eq. (\ref{wc(k)}) and marked by the two vertical dashed lines on the left for $k=-1.5$ and $k=-1.1$. The unconstrained right edge $l_k^{\rm uc}$, for these two values of $k$, are also marked by two vertical dashed lines on the right.}
    \label{fig:km1p5bc}
\end{figure}
This implies 
\begin{equation}
\begin{split}
    \Big[\left(\gamma_k \big(1-2q_k(w)\big)+z\right)(g_k(w)-z) + \frac{\mathcal{B}_k}{\mathcal{A}_k} \frac{2\gamma_k g_k(w)}{(g_k(w)(g_k(w)-1))^{\gamma_k}}\Big] =  z(1-z) \;.
\end{split}
\end{equation}
Matching the powers of $z$ on both sides gives two relations
\begin{equation}\label{qkw}
    q_k(w) = \frac{\gamma_k+1 -g_k(w)}{2\gamma_k},
\end{equation}
and
\begin{equation}\label{dkw}
    \frac{\mathcal{B}_k}{\mathcal{A}_k} = -\frac{D_k^*(w)}{\mathcal{A}_k}= \frac{g_k(w)^{\gamma_k}(g_k(w)-1)^{\gamma_k+1}}{|1+k|} \;,
\end{equation}
where we used ${\cal B}_k = -D_k^*(w)$ from Eq. (\ref{constants_3}). Solving these Eqs.~\eqref{qkw} and~\eqref{dkw} one can get $D^*_k$ and $\bar{l}_k(w)$ in terms of $l_k(w)$
\begin{equation}\label{lwtkm1p5}
    \bar{l}_k(w) = \frac{2 w + (k+1) l_{k}(w)}{k+3},
\end{equation}
\begin{equation}\label{dwkm1p5}
D^*_k(w) = \frac{\left(l_{k}(w)-w\right)\left(w+l_{k}(w)\right)^{\frac{k+1}{2}}}{\left|k\right|\left(k+3\right)} \left(\frac{(k+1)\left(w-l_{k}(w)\right)}{k+3}\right)^{\frac{k+1}{2}}.
\end{equation}
The only remaining constant $l_k(w)$ is finally determined from the normalization condition Eq.~\eqref{norm_cond}. The scaled bulk density is then given by
\begin{equation}
    \phi_k(z, w) = -A_k |k| \mathcal{A}_k\frac{z^{\gamma_k}\left(1-z\right)^{\gamma_k+1}}{g_k(w)-z} \;,
\end{equation}
which in terms of the original coordinates reads
\begin{equation}\label{dens:klm1}
    \rho^*_b(y,w) = 
    A_k  \left(l_{k}(w)+y\right)^{\gamma_k} \frac{(\bar{l}_k(w)-y)^{\gamma_k+1}}{(w-y)} \quad {\rm where} \quad \gamma_k = \frac{k+1}{2} \;,
\end{equation}
and the constant $A_k$ is given in Eq.~\eqref{A_1k}.
\begin{figure}[t]
    \centering
    \includegraphics[scale=0.6]{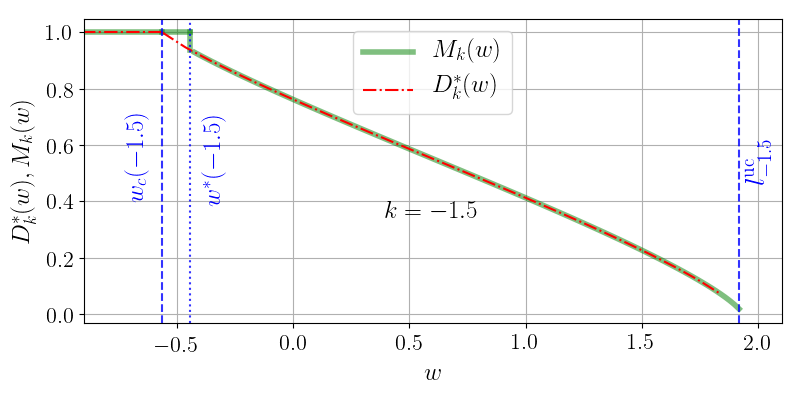}
    \caption{\textit{Regime $3$}: Simultaneous plots (i) of the amplitude $D^*_k(w)$ of the delta function, associated with the density $\rho^I(y,w)$ in Eq. (\ref{rhoI}) (dashed red line) and (ii) of the order parameter $M_k(w)$ defined in Eq. (\ref{def_Mk}) (solid green line), as a function of $w$ for fixed $k=-1.5$. We see that $D_k^*(w)$ increases with decreasing $w$ and approaches to $1$ as $w \to w_c(k)$, while the order parameter $M_k(w)$ coincides with $D_k^*(w)$ for $w>w^*(k)$ but jumps to $1$ at $w=w^*(k)$. This jump in $M_k(w)$ at $w=w^*(k)$ demonstrates a first-order phase transition.}
    \label{fig:dwvsw}
\end{figure}
Inserting this density~\eqref{dens:klm1} in the normalization condition Eq.~\eqref{norm_cond} and using the expression for $D_k^*(w)$ in (\ref{dwkm1p5}) we find that $l_k(w)$ satisfies the equation
\begin{equation}\label{l_{k}(w)-k<-1}
\frac{g_k(w)(2\gamma_k+1)-(\gamma_k+1)}{\gamma_k B(\gamma_k+1, \gamma_k+1)^{-\alpha_k}}\left(I\left(g_k(w), \gamma_k, \gamma_k+1\right)+\frac{d_k(w)}{A_k}\right)^{-\alpha_k} = \frac{w}{l_k^{\rm uc}},
\end{equation}
where we recall that $g_k(w) = {L_k(w)}/{\tilde{L}_k(w)}$ with $L_k(w) = w+l_k(w)$ and $\tilde L_k(w) = \bar{l}_l(w)+l_k(w)$. Here 
\begin{align}\label{Id}
I(g_k(w), \gamma_k, \gamma_k+1)=\int_0^1 \frac{z^{\gamma_k}(1-z)^{\gamma_k+1}}{g_k(w)-z}\,dz \quad {\rm and} \quad d_k(w) = \frac{g_k(w)^{\gamma_k}\left(g_k(w)-1\right)^{\gamma_k+1}}{k(k+1)} \;.
\end{align}
For a given $w$ and $k$, we solve this equation numerically to get $g_k(w)$. Using $g_k(w)= (w+l_k(w))/\tilde L_k(w)$ where 
\begin{equation}
\tilde{L}_k(w) = \bar{l}_k(w)+l_k(w) = \frac{2\gamma_kw}{(2 \gamma_k+1)g_k(w)-(1+\gamma_k)} \;,
\label{Lktil}
\end{equation}
we then finally fix the only unknown $l_k(w)$. We then have the full analytical expression of the bulk density $\rho_b^*(y,w)$ in Eq. (\ref{dens:klm1}) and the weight of the delta function $D_k^*(w)$ in Eq.~(\ref{dwkm1p5}). The numerical results from MC simulation for the density shown in Fig. \ref{fig:km1p5} are in good agreement with these analytical predictions. A slight complication arises in the MC simulation as it turns out that the particles have extremely small fluctuations and hence are confined to a small region around there mean position. So to get a better thermal average of density we need to go to larger $N$ which is computationally costly. We bypassed this issue by considering comparatively high temperature $T=1000$ but still satisfying the constraint $\beta N^{2\alpha_k}\gg1$. At such high temperature particles fluctuate more leading to a smoother density profile for the chosen values of $N$. We notice that the numerical densities match better with the expression Eq.~\eqref{dens:klm1} for larger values of~$N$.

Let us first analyse the limit $w \to l_k^{\rm uc}$ from the left. In this limit, the right hand side of Eq. (\ref{l_{k}(w)-k<-1}) approaches to $1$. Consequently, one can show, by analysing the left hand side of Eq. (\ref{l_{k}(w)-k<-1}) that $g_k(w) =  {L_k(w)}/{\tilde{L}_k(w)} \to 1$ in that limit. Consequently, $L_k(w) = w+l_k(w)$ approaches $\tilde L_k(w) = l_k(w) + \bar{l}_k(w)$. Hence, $\bar l_k(w) \to w$ indicating that the hole disappears in this limit. In addition, from Eq. (\ref{Lktil}), it follows that $l_k(w) \to w$ and the support length $\tilde L_k(w) \to 2 l_k^{\rm uc}$. In addition, the weight of the delta-peak in Eq.~(\ref{dwkm1p5}) vanishes in this limit. We thus fully recover the 'U-shaped' unconstrained density, as in the first row of Fig. \ref{fig:allkdens}. 

\begin{figure}[t]
  \centering
     \includegraphics[scale=0.5]{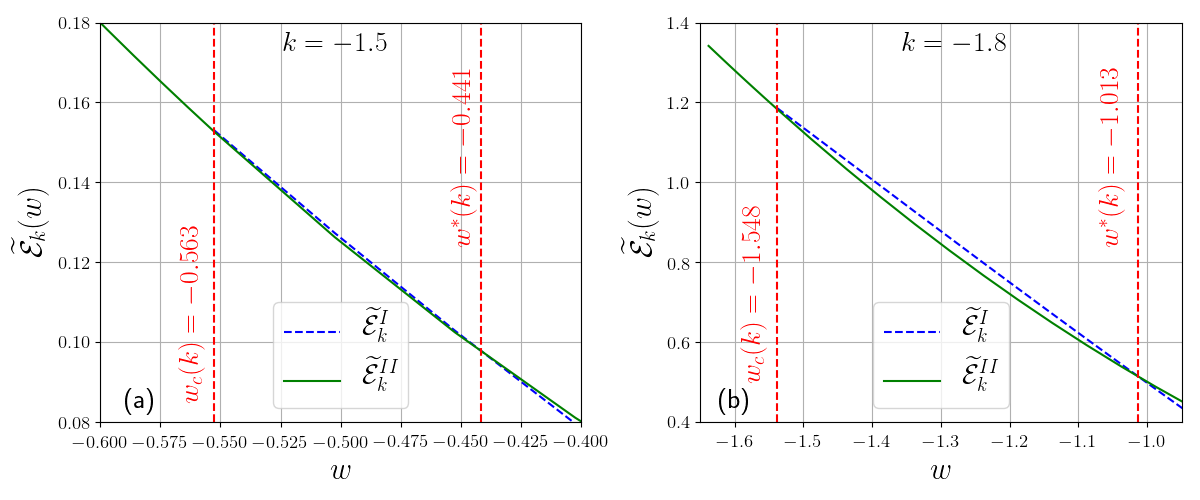}
     \caption{Plot of the energy $\tilde{\cal E}_k^I(w)$ and $\tilde{\cal E}_k^{II}(w) = w^2/2$ vs. $w$ for two different values of $k$: for $k=-1.5$ (panel (a)) and $k=-1.8$ (panel (b)). The values of $w^*(k)$ and $w_c(k)$ are marked by vertical dashed lines. In the range $w_c(k)< w < w^*(k)$ the energy $\tilde{\cal E}_k^{I}(w) >  \tilde{\cal E}_k^{II}(w)$, showing that the configuration with density $\rho_k^I(y,w)$ is metastable.}
     \label{fig:compar}
\end{figure}
Now consider pushing the position of the wall $w$ further to the left. As $w$ decreases, more and more particles get transferred from the extended bulk to the delta-peak. 
As a result, the support of the bulk density $\tilde L_k(w)$ reduces monotonically with decreasing $w$ (see Fig. \ref{fig:km1p5bc}) and the weight of the delta-peak $D_k^*(w)$ increases monotonically with decreasing $w$ (see Fig. \ref{fig:dwvsw}). It turns out that there is a critical value $w_c(k)$ at which $D_k^*(w)$ hits $1$ and simultaneously $\tilde L_k(w)$ hits $0$. At $w=w_c(k)$, there are no particles left in the extended bulk and the wall absorbs all the particles. If $w$ is decreased below $w_c(k)$, all the particles are still at the wall and the density remains a delta-function at the wall, i.e., 
\begin{equation}\label{delta_density}
    \rho^*_k(y,w) = \delta(w-y) \quad \text{for} \quad w\leq w_c(k) \;.
\end{equation}

To determine the critical value $w_c(k)$, we first note that the support length $\tilde L_k(w) = l_k(w)+\bar{l}_k(w) = 0$ at $w=w_c(k)$. Using $\bar{l}_k(w_c) = - l_k(w_c)$ in Eq. (\ref{lwtkm1p5}) gives $l_k(w_c) = - w_c/(k+2)$. Substituting this value in the expression of $D_k^*(w)$ in Eq. (\ref{dwkm1p5}) and setting $D_k^*(w_c)=1$ 
gives
\begin{equation}\label{wc(k)}
    w_c(k) = \frac{(k+2) \left| k (k+1)\right| ^{\frac{1}{k+2}}}{k+1} \;.
\end{equation} 
Since $-2<k<-1$, $w_c(k)<0$. Note that in the limit $k \to -1$, $w_c(k) \to -1$, which is indeed the left edge of the unconstrained scaled density \cite{dhar2017exact,dhar2018extreme}.   

\vspace*{0.5cm}
\noindent{\it Metastability and first-order phase transition.}
So far, we have assumed that the optimal density profile for $-2<k < -1$ is given by the ansatz in Eq. (\ref{dens:ansatz}) which consists of a disjoint bulk part and a 
delta-peak at the wall, separated by a hole in between. We will denote this solution by the superscript $I$ and it reads
\begin{eqnarray}\label{rhoI}
\rho^{I}_k(y, w)=\rho^*_b(y, w)\mathbb{I}[-l_k(w)\leq y \leq \bar{l}_k(w)]+ D^*_k(w) \delta (w-y) \;.
\end{eqnarray}
\begin{figure}[t]
\centering
\includegraphics[width = 0.9\linewidth]{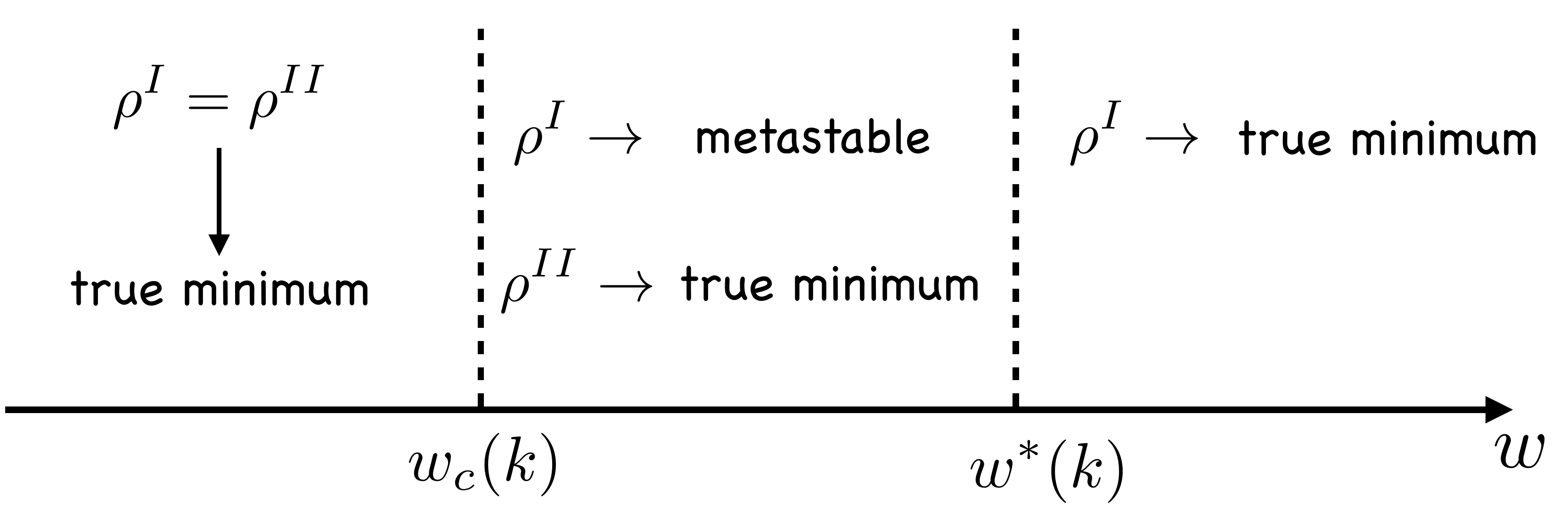}
\caption{The optimal density is one of two different types $\rho^{I}_k(y, w)$ and $\rho^{II}_k(y, w)$ defined respectively in Eqs. (\ref{rhoI}) and (\ref{rhoII}). For $w>w^*(k)$,  $\rho^{I}_k(y, w)$ is the optimal density. For $w_c(k)<w<w^*(k)$, the density  $\rho^{I}_k(y, w)$ becomes metastable, while $\rho^{II}_k(y, w)$ represents the true minimum. Finally, for $w<w_c(k)$, the solutions $\rho^{I}_k(y, w)$ and  $\rho^{II}_k(y, w)$ merge with each other.}\label{Fig_meta}
\end{figure}
We have seen that for $w\leq w_c(k)$ this density becomes a pure delta-peak located at $w$, with $w_c(k)$ given in Eq. (\ref{wc(k)}). This suggests that there could
be a candidate configuration for a minimum energy, denoted by a superscript $II$, which consists of a pure delta-function at $w$ {\it for any $w$}, and not just for $w \leq w_c(k)$. It reads
\begin{eqnarray}\label{rhoII}
\rho^{II}_k(y, w) = \delta(y-w) \;.
 \end{eqnarray}
These two candidate configurations $\rho^{I}_k(y, w)$ and $\rho^{II}_k(y, w)$ merge for $w \leq w_c(k)$. Hence for $w > w_c(k)$, one naturally wonders which one of these two configurations $\rho^{I}_k(y,w)$ and $\rho^{II}_k(y,w)$ has the lower energy. To answer this question, we need to evaluate the energy in Eq. (\ref{scaled-field}) associated to these two density profiles and compare them for $w > w_c(k)$. Let us denote the two energies by $\tilde {\cal E}_k^{I}(w)$ and $\tilde {\cal E}_k^{II}(w)$ respectively. The energy $\tilde {\cal E}_k^{II}(w)$ is very simple and is given by just  $\tilde {\cal E}_k^{II}(w)=w^2/2$. In contrast, the energy $\tilde {\cal E}_k^{I}(w)$ has to be evaluated from Eq. (\ref{ener_func_3}) with the substitution $\rho_b(y,w)=\rho_b^*(y,w)$ as given explicitly in Eq. (\ref{dens:klm1}) and $D_k(w) = D_k^*(w)$ as given in Eq. (\ref{dwkm1p5}).
It is a bit hard to obtain an explicit formula for $\tilde {\cal E}_k^{I}(w)$ but it can be evaluated numerically very accurately. The results are shown in Fig. \ref{fig:compar} for two different values of $k$. 
Surprisingly, it turns out that there is yet another critical value $w^*(k)>w_c(k)$ such that
\begin{align} \label{comparison_E}
 \tilde {\cal E}_k^{I}(w) < \tilde {\cal E}_k^{II}(w) = \frac{w^2}{2} &\quad \text{     when     } \quad w>w^*(k)\\
\tilde {\cal E}_k^{I}(w) > \tilde {\cal E}_k^{II}(w) = \frac{w^2}{2} & \quad \text{     when     } \quad w_c(k)<w<w^*(k) \;.
\end{align}
Thus for $w> w^*(k)$, the density $\rho^{I}_k(y, w)$ is the true optimal solution, while in the intermediate range $w_c(k)<w<w^*(k)$ the solution $\rho^{II}_k(y, w)$ (pure delta peak) turns out to be the true minimum. Thus for $w_c(k)<w<w^*(k)$ the solution $\rho^{I}_k(y, w)$ corresponds to a ``metastable'' minimum. Numerically we find that, in this intermediate region, the two energies $\tilde {\cal E}_k^{I}(w)$ and $\tilde {\cal E}_k^{II}(w)$ are very close to each other (see Fig. \ref{fig:compar}). Hence to summarise, the true optimal density profile is given by
\begin{eqnarray} \label{summary_rho}
\rho_k^*(y,w) = 
\begin{cases}
&\rho^{I}_k(y, w) \quad {\rm for} \quad w > w^*(k) \;,\\
& \\
&\rho^{II}_k(y, w) \quad {\rm for} \quad w<w^*(k) \;.
\end{cases}
\end{eqnarray}
\begin{figure}[t]
  \centering
     \includegraphics[scale=0.5]{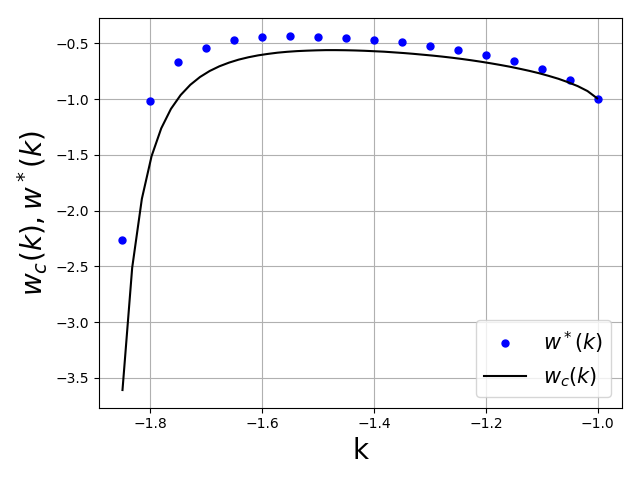}
     \caption{{Plot of the critical wall positions $w_c(k)$ and $w^*(k)$ as functions of $k$ in regime 3 $-2<k\leq -1$. $w_c(k)$ (black solid line) is given by Eq.~\eqref{wc(k)} and $w^*(k)$ (blue dots) is found numerically from the crossover location between the energies $\tilde {\cal E}^I_k(w)$ and $\tilde {\cal E}^{II}_k(w)$ as shown in Fig~\ref{fig:compar}. We find that $w_c(k) \leq w^*(k)$ for all $-2<k\leq-1$ with $w_c(k)=w^*(k)$ only for $k=-1$.}}
     \label{fig:wcws}
\end{figure}
These behaviours are summarised in Fig.~\ref{Fig_meta}. Thus we see that the system undergoes a first-order phase transition at $w=w^*(k)$ where the true minimum density changes abruptly from $\rho^I$ to $\rho^{II}$ as $w$ crosses $w^*(k)$ from above. A manifestation of this first-order phase transition can be observed in the order parameter defined as the amplitude of the delta peak in the true optimal solution $\rho_k^*(y,w)$
\begin{eqnarray}\label{def_Mk}
M_k(w) = 
\begin{cases}
& D_k^*(w) \quad, \quad w > w^*(k) \\
& 1 \quad\quad\quad, \quad w < w^*(k) \;.
\end{cases}
\end{eqnarray}
For $w>w^*(k)$ it is given by $D_k^*(w)$ in Eq. (\ref{dwkm1p5}) associated with the density $\rho^{I}$. When $w$ goes below $w^*(k)$ this amplitude undergoes a jump to $1$ corresponding to the full delta function $\rho^{II}$ in Eq. (\ref{rhoII}). In Fig. \ref{fig:dwvsw} we have plotted both $D_k^*(w)$ associated with the density $\rho^I$ and the true order parameter $M_k(w)$ given in Eq. (\ref{def_Mk}). Thus $M_k(w)$ undergoes a jump at $w=w^*(k)$, demonstrating a first-order phase transition. 

\begin{figure}[t]
  \centering
     \includegraphics[scale=0.6]{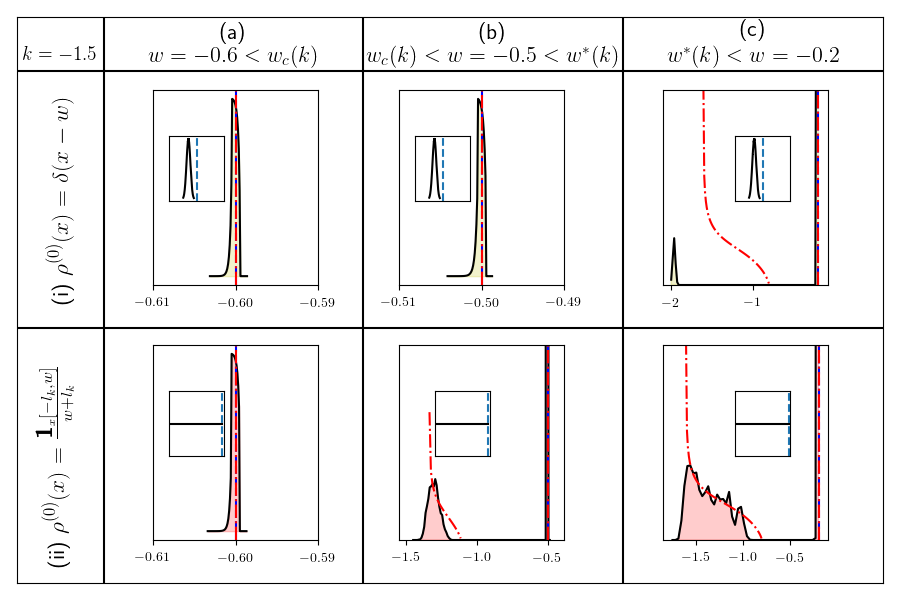}
     \caption{In this figure we study the metastability of the extended profile in region $w_c(k)< w< w^*(k)$ for $k=-1.5$. For this value of $k$, $w_c(k)=-0.563$ and $w^*(k)=-0.441$. The figure is divided into three columns  corresponding to the three regimes (a) $w=-0.6<w_c(k)$, (b) $w_c(k)< w=-0.5< w^*(k)$ and (c) $w=-0.2>w^*(k)$. For each column, the insets of the top row and the bottom row indicate two different initial conditions (a delta peak and a flat density) while the main figures show the final configuration after a large number of MC steps. In columns (a) and (c) we see see that the final configurations in the top row and in the bottom row are qualitatively similar, indicating the irrelevance of the initial conditions. In contrast, in column (b) the final configurations in the top and in the bottom row corresponding to two different initial conditions seem to lead to different final configurations, within the time scale of the simulation. This dependence on the initial condition is a signature of metastability in region (b).}
     \label{fig:frozen}
\end{figure}


In order to check this scenario numerically, we have performed MC simulations. We have first determined $w^*(k)$ numerically by evaluating the energy of the
solutions $\rho^{I}$ and $\rho^{II}$. In Fig.~\ref{fig:wcws} we plot $w^*(k)$ (numerical) and $w_c(k)$ (analytical from Eq. (\ref{wc(k)})) as a function of $k$ for $-2<k\leq -1$. We observe that the difference between the two is rather small but clearly $w_c(k)<w^*(k)$. In fact the difference between them 
vanishes as $k \to -1$ (see Fig.~\ref{fig:wcws}). This is expected because we know from the exact solution of the case $k=-1$ ($1d$OCP) that $\rho^{I}_k(y, w)$ is the exact optimal solution for all $w$ \cite{dhar2017exact,dhar2018extreme}. To test the metastability in the intermediate regime $w_c(k)<w<w^*(k)$, 
we consider three different wall positions (a) $w<w_c(k)$, (b) $w_c(k)< w< w^*(k)$ and (c) $w>w^*(k)$ for $k=-1.5$. For each wall position we study two distinct initial conditions: (i) delta function and (ii) a uniform density profile and observe the steady state profiles. In Fig.~\ref{fig:frozen}, we find that for cases (a) and (c) the steady state profile is independent of the initial conditions and converges respectively to $\rho^{II}_k(y, w)$ and $\rho^{I}_k(y, w)$. On the other hand in case (b) the late time profile (within the time scale of the simulation) depends on the initial conditions -- a typical hallmark of metastability. More precisely, if one starts with a delta function profile, the late time configuration remains a delta function whereas if the initial profile is uniform then the late time profile seems to stay closer to $\rho^{I}_k(y, w)$, within the time scale of the simulation. This picture is thus fully consistent with our discussion that for $w_c(k)<w<w^*(k)$ the density profile $\rho^{I}_k(y, w)$ is metastable and the true minimum is given by $\rho^{II}_k(y, w)$.

\section{Conclusions}
\label{conclusion}

In this paper, we have studied the average density of a harmonically confined Riesz gas of $N$ particles for large $N$ in the presence of a hard wall located at $W$. 
In this Riesz gas, the particles repel each other via a pairwise interaction that behaves as $|x_i - x_j|^{-k}$ for $k>-2$, with $x_i$ denoting the position of the $i^{\rm th}$ particle. Our goal was to study how the equilibrium density of the gas, in the large $N$ limit, gets modified in the presence of the wall. We have computed exactly this average density in the limit of large $N$. This density can be classified into three different regimes of $k$, as depicted in Figs. \ref{Fig_k} and \ref{fig:allkwdens}. For $k \geq 1$, where the interactions are effectively short-ranged, the appropriately scaled density has a finite support over $[-l_k(w),w]$ where $w$ is the scaled position of the wall. While the density vanishes at the left edge of the support, it approaches a nonzero constant at the right edge $w$. For $-1<k<1$, where the interactions are weakly long-ranged, we find that the scaled density is again supported over $[-l_k(w),w]$. While it vanishes at the left edge of the support, it diverges at the right edge $w$ algebraically with an exponent $(k-1)/2$. For $-2<k\leq -1$, the interactions are strongly long-ranged that leads to a rather exotic density profile: here the density has an extended bulk part and a delta-peak at the wall and they are separated by a hole for $-2<k<-1$. Interestingly, we find that there is a first-order phase transition at a critical value $w=w^*(k)$ such that for $w<w^*(k)$ the optimal solution changes its nature. It consists only of a delta peak at the wall, i.e., the wall essentially absorbs all the particles and there is no extended part. The amplitude of the delta-peak $M_k(w)$ plays the role of an order parameter which undergoes a jump to a value $1$ as $w$ is decreased through $w^*(k)$.

As mentioned in the introduction, the cumulative distribution of the position of the right-most particle $x_{\max}$ is closely related to this density in the presence of a wall in the large $N$ limit [see Eq. (\ref{EVS})]. Therefore the results obtained here will be an essential ingredient to compute the probability of large deviations of $x_{\max}$ for any $k>-2$. Indeed, this large deviation behavior of $x_{\max}$ has so far been computed only for two specific values of $k$, namely $k \to 0$ limit \cite{dean2006large, dean2008extreme} and $k=-1$ \cite{dhar2017exact,dhar2018extreme}. In the former case, this is also the large deviation of the top eigenvalue of a Gaussian random matrix. In that context, it is known that when the wall hits the right edge of the unconstrained density, it is accompanied by a third-order phase transition, where the third derivative of the large deviation function has a discontinuity \cite{majumdar2014top}. Interestingly, a similar third-order phase transition occurs also for $k=-1$~\cite{dhar2017exact,dhar2018extreme}. It will be interesting to investigate whether this transition remains third-order for other values of $k$ \cite{jit2021evs}. Furthermore, in the regime $-2<k<-1$ we have seen that there is a first-order transition in the density of the Riesz gas at a critical value $w=w^*(k)$. It will be interesting to study the implications of this first-order transition for the large deviation behavior of $x_{\max}$ in this strongly long-ranged regime.

\ack
We thank Abhishek Dhar for helpful discussions and suggestions. M.~K. would like to acknowledge support from the project 6004-1 of the Indo-French Centre for the Promotion of Advanced Research (IFCPAR), Ramanujan Fellowship (SB/S2/RJN-114/2016), SERB Early Career Research Award (ECR/2018/002085) and SERB Matrics Grant (MTR/2019/001101) from the Science and Engineering Research Board (SERB), Department of Science and Technology, Government of India. M.~K.  and A.~K. acknowledge support from the Department of Atomic Energy, Government of India, under Project No. RTI4001. A.~K. acknowledges support from DST, Government of India 
grant under project No. ECR/2017/000634. S.~N.~M. thanks the warm hospitality of the Weizmann Institute as a visiting Weston fellow and ICTS where this work was completed.  D. M. acknowledges the support of the Center of Scientific Excellence at the Weizmann Institute of Science.
\appendix

\section{}
In this Appendix, we detail our derivation of the constrained scaled density for $-1<k<1$ in Eq. (\ref{phikl11}), and for 
and for $-2<k\leq-1$ in Eq. (\ref{phi:klm1}).

\subsection{Regime $2: -1<k<1$}
\label{apndkl1}
We start with the solution of the Sonin equation \eqref{kl1 int eq 1}, as given in Eqs.~\eqref{sol:sonin} and~\eqref{sol:sonin1}. For convenience we rewrite here the particular solution $u_k(z)$ in Eq.~\eqref{sol:sonin1} 
\begin{equation}\label{part_kl1_1}
u_k(z)=\frac{2 A_k |k|z^{\frac{k-1}{2}}}{B\left(\frac{k+1}{2},\frac{k+1}{2}\right)} \frac{\partial }{\partial z} \left(\int_z^{1} t^{-k} (t-z)^{\frac{k+1}{2}} \frac{\partial }{\partial t }\int_0^t h(y)y^{\frac{k+1}{2}} (t-y)^{\frac{k-1}{2}} \, dy \, dt\right) \;,
\end{equation}
where $h_k(z) =\mathcal{A}_k(z-q_k(w))$. This Eq.~\eqref{part_kl1_1} can be written as
\begin{equation}
\label{uksim}
u_k(z)=\frac{A_k |k|}{B\left(\frac{k+1}{2},\frac{k+1}{2}\right)}z^{\frac{k-1}{2}} \frac{\partial }{\partial z} I_2(z,k)
\end{equation}
where
\begin{equation}
\label{i2}
 I_2(z, k) = \int_z^{1} dt \, t^{-k} (t-z)^{\frac{k+1}{2}} \frac{\partial }{\partial t }I_1(t, k)
\end{equation}
\begin{equation}
\label{i1}
 I_1(t,k)= \int_0^t \, dy \, h_k(y)y^{\frac{k+1}{2}} (t-y)^{\frac{k-1}{2}} \;.
\end{equation}
The integral $I_1(t,k)$ in Eq.~\eqref{i1} can be computed explicitly and we get
\begin{equation}
\begin{split}
    I_1(t, k) = \frac{\mathcal{A}_k t^{1+k}}{2} B\left(\frac{k+1}{2}, \frac{k+1}{2}\right)\left( t \left(\frac{k+3}{2(k+2)}\right) -  q_k(w)\right).
\end{split}
\label{part_kl1_2}
\end{equation}
Taking a derivative of the Eq.~\eqref{part_kl1_2} with respect to $t$, we get
\begin{equation}
\begin{split}
    \frac{\partial}{\partial t} I_1(t, k) = B\left(\frac{k+1}{2}, \frac{k+1}{2}\right)\Bigg(\mathcal{A}_k t^{k+1} \left(\frac{k+3}{4}\right) -  \frac{\mathcal{A}_k q_k(w)(k+1)}{2} t^k\Bigg).
\end{split}
\end{equation}
Substituting this result in Eq. (\ref{i2}), the integral $I_2(z,k)$ reads
\begin{equation}
\begin{split}
    I_2(z,k)=B\left(\frac{k+1}{2}, \frac{k+1}{2}\right)\Bigg(&\int_z^1 dt \, (t-z)^{\frac{k+1}{2}} \frac{\mathcal{A}_k(k+3)}{4}t\\&- \int_z^1 dt \, (t-z)^{\frac{k+1}{2}} \frac{\mathcal{A}_kq_k(w)(k+1)}{2}\Bigg).
\end{split}
\end{equation}
Now taking a derivative with respect to $z$ gives
\begin{equation}
\begin{split}
    \frac{\partial}{\partial z}I_2(z,k) = B\left(\frac{k+1}{2}, \frac{k+1}{2}\right)\mathcal{A}_k(1-z)^{\frac{k+1}{2}}\Bigg(\frac{q_k(w)(k+1)}{2} -\frac{(1+k+2z)}{4} \Bigg) \;.
\end{split}
\end{equation}
We then finally get from Eq. (\ref{uksim})
\begin{equation}
\begin{split}
    u_k(z) = A_k |k|z^{\frac{k-1}{2}}(1-z)^{\frac{k+1}{2}} \mathcal{A}_k \Bigg(q_k(w)(k+1) - \frac{1+k+2z}{2}\Bigg) \;.
\end{split}
\end{equation}
In terms of $\gamma_k = \frac{k+1}{2}$ it reads
\begin{equation}
\begin{split}
    u_k(z) = -A_k |k|\mathcal{A}_k z^{\gamma_k-1}(1-z)^{\gamma_k}\big(z-\gamma_k(2q_k(w)-1)\big) \;.
\end{split}
\end{equation}
Substituting this in Eq.~\eqref{sol:sonin} gives
\begin{equation}
\begin{split}
    \phi_k(z, w) = \big(z(1-z)\big)^{\gamma_k-1}\Bigg( C_0 - A_k |k|  \mathcal{A}_k (1-z)\left( z - \gamma_k \big(2q_k(w)-1\big)\right) 
     \Bigg) \;,
\end{split}
\end{equation}
which is indeed Eq. (\ref{phikl11}) in the main text. 

\subsection{Regime $3: -2<k\leq-1$}
\label{apndklm1}
Equation~\eqref{longklm1:integraleqn2} in the main text reads explicitly
\begin{equation}\label{longklm1:integraleqn4}
    \int_0^1 dz' \frac{sgn(z'-z)}{|z'-z|^{k+1}} \phi_k(z', w) = \mathcal{A}_k(z-q_k(w)) + \mathcal{B}_k(g_k(w)-z)^{-(k+1)}
\end{equation}
where $\mathcal{A}_k = - \frac{\tilde{L}_k(w)^{k+2}}{|k|}$, $g_k(w) = \frac{w+l_w}{\tilde{l}_w + l_w}$, $q_k(w) = \frac{l_k(w)}{\tilde{L}_k(w)}$ and $\mathcal{B}_k = -D^*_k(w)$.
The solution of the Sonin equation~\eqref{longklm1:integraleqn4} is given by
\begin{equation}\label{sol:sonin9}
    \phi_k(z, w) = C_0\big(z(1-z)\big)^{\frac{k-1}{2}} + u_k(z)
\end{equation}
with the particular solution $u_k(z)$ given by
\begin{equation}\label{sol:sonin6}
u_k(z)=\frac{2A_k |k|z^{\frac{k-1}{2}}}{B\left(\frac{k+1}{2},\frac{k+1}{2}\right)} \frac{\partial }{\partial z} \left(\int_z^{1} dt \, t^{-k} (t-z)^{\frac{k+1}{2}} \frac{\partial }{\partial t }\int_0^t \, dy \, h_k(y)y^{\frac{k+1}{2}} (t-y)^{\frac{k-1}{2}}  \right)
\end{equation}
where $h_k(z) =\mathcal{A}_k(z-q_k(w)) + \mathcal{B}_k(g_k(w)-z)^{-(k+1)}$. This can be written as
\begin{equation}\label{sol:sonin10}
u_k(z)=\frac{A_k |k|}{B\left(\frac{k+1}{2},\frac{k+1}{2}\right)}z^{\frac{k-1}{2}} \frac{\partial }{\partial z} I_2(z,k)
\end{equation}
where
\begin{equation} \label{A16}
 I_2(z, k) = \left(\int_z^{1} dt \, t^{-k} (t-z)^{\frac{k+1}{2}} \frac{\partial }{\partial t }I_1(t, k)\right)
\end{equation}
\begin{equation} \label{A17}
 I_1(t,k)= \int_0^t \, dy \, h_k(y)y^{\frac{k+1}{2}} (t-y)^{\frac{k-1}{2}} \;.
\end{equation}
The integral $I_1(t, k)$ in Eq.~\eqref{A17} becomes
\begin{equation}
\begin{split}
    I_1(t, k) = \frac{\mathcal{A}_k t^{1+k}}{2} &B\left(\frac{k+1}{2}, \frac{k+1}{2}\right)\left( t \left(\frac{k+3}{2(k+2)}\right) -  q_k(w)\right) \\&+ \mathcal{B}_k \int_0^1 dr\,  \frac{r^{\frac{k+1}{2}}(1-r)^{\frac{k-1}{2}}}{(\frac{g_k(w)}{t}-r)^{k+1}} \;.
\end{split}
\label{sol:sonin7}
\end{equation}
The integral in the second term in Eq~\eqref{sol:sonin7} can be done by a change of variable
\begin{equation}\label{variablechange}
s = \frac{r(\frac{g_k(w)}{t}-1)}{\frac{g_k(w)}{t}-r} \;,
\end{equation}
where we have assumed that $\frac{g_k(w)}{t} \geq 1$ (which can be verified a posteriori). The integral in Eq~\eqref{sol:sonin7} then becomes
\begin{equation}
    \int_0^1 dr\, \frac{r^{\frac{k+1}{2}}(1-r)^{\frac{k-1}{2}}}{(\frac{g_k(w)}{t}-r)^{k+1}} = \frac{1}{2}\left(\frac{t}{g_k(w)}\right)^{k+2}B\left(\frac{k+1}{2}, \frac{k+1}{2}\right)\left(1-\frac{t}{g_k(w)}\right)^{-\frac{k+3}{2}} \;.
\end{equation}
Taking a derivative of Eq.~\eqref{sol:sonin7} with respect to $t$, we get
\begin{equation}
\begin{split}
    \frac{\partial}{\partial t} I_1(t, k) = B\left(\frac{k+1}{2}, \frac{k+1}{2}\right)\Bigg(& \mathcal{A}_k t^{k+1} \left(\frac{k+3}{4}\right) -  \frac{\mathcal{A}_k q_k(w)(k+1)}{2} t^k \\&+ \frac{\mathcal{B}_k(k+1)}{2 g_k(w)} \left(\frac{t}{g_k(w)}\right)^{k} \left(1 - \frac{t}{g_k(w)}\right)^{-\frac{k+3}{2}} \Bigg)
\end{split}
\end{equation}
Substituting this in Eq. (\ref{A16}) we get
\begin{equation}
\begin{split}
    I_2(z,k)&=B\left(\frac{k+1}{2}, \frac{k+1}{2}\right) \int_z^1 dt \, (t-z)^{\frac{k+1}{2}} \frac{\mathcal{A}_k(k+3)}{4}t  \\&- B\left(\frac{k+1}{2}, \frac{k+1}{2}\right)\int_z^1 dt \, (t-z)^{\frac{k+1}{2}} \frac{\mathcal{A}_kq_k(w)(k+1)}{2}  \\&  +B\left(\frac{k+1}{2}, \frac{k+1}{2}\right)\int_z^1 dt \, (t-z)^{\frac{k+1}{2}}\frac{\mathcal{B}_k(k+1)}{2 g_k(w)^{k+1}}\left(1 - \frac{t}{g_k(w)}\right)^{-\frac{k+3}{2}}.
\end{split}
\end{equation}
This finally gives, from Eq. (\ref{sol:sonin10})
\begin{equation}
\begin{split}
    \frac{\partial}{\partial z}I_2(z,k) = B\left(\frac{k+1}{2}, \frac{k+1}{2}\right)\mathcal{A}_k\frac{(1-z)^{\frac{k+1}{2}}}{g_k(w)-z}\Bigg(\left(\frac{(k+1)(g_k(w)-z)q_k(w)}{2}\right)&\\ -\left(\frac{(1+k+2z)(g_w-z)}{4}-\frac{\mathcal{B}_k g_k(w)(k+1)}{2\mathcal{A}_k(g_k(w)(g_k(w)-1))^{\frac{k+1}{2}}}\right) \Bigg).
\end{split}
\end{equation}
This finally gives
\begin{equation}
\begin{split}
    u_k(z) = A_k |k| \frac{z^{\frac{k-1}{2}}(1-z)^{\frac{k+1}{2}}}{g_k(w)-z} \mathcal{A}_k \Bigg(&\left(q_k(w)(k+1)- \frac{1+k+2z}{2}\right)(g_k(w)-z) \\&-\frac{\mathcal{B}_k}{\mathcal{A}_k} \frac{g_k(w)(k+1)}{(g_k(w)(g_k(w)-1))^{\frac{k+1}{2}}}\Bigg) \;.
\end{split}
\end{equation}
In terms of $\gamma_k = \frac{k+1}{2}$ it reads
\begin{equation}
\begin{split}
    u_k(z) = -A_k |k| \frac{z^{\gamma_k-1}(1-z)^{\gamma_k}}{g_k(w)-z} \mathcal{A}_k \Bigg(&\big(\gamma_k(1-2q_k(w))+z\big)(g_k(w)-z) +\\& \frac{\mathcal{B}_k}{\mathcal{A}_k} \frac{2\gamma_kg_k(w)}{(g_k(w)(g_k(w)-1))^{\gamma_k}}\Bigg) \;.
\end{split}
\end{equation}
Substituting this in Eq.~\eqref{sol:sonin9} gives
\begin{equation}
\begin{split}
    \phi_k(z, w) = \big(z(1-z)\big)^{\gamma_k-1}\Bigg[ C_o& - A_k |k| (1-z) \mathcal{A}_k \left(\gamma_k \big(1-2q_k(w)\big)+z\right)\\& - A_k |k| \frac{1-z}{g_k(w)-z} \mathcal{B}_k \frac{2\gamma_k g_k(w)}{(g_k(w)(g_k(w)-1))^{\gamma_k}}
     \Bigg] \;,
\end{split}
\end{equation}
which is indeed Eq. (\ref{phi:klm1}) in the main text.

\begin{center}
\line(1,0){250}
\end{center}
\clearpage
\section*{References}

\end{document}